\documentclass[preprint]{aastex}

\def\cmodel{{\tt cmodel\_counts}}
\def\model{{\tt counts\_model}}
\def\exp{{\tt counts\_exp}}
\def\dev{{\tt counts\_dev}}
\def\petro{{\tt petrocounts}}
\def\psf{{\tt psfcounts}}
\def\photo{{\tt photo}}
\def\uprime{u}
\def\gprime{g}
\def\rprime{r}
\def\iprime{i}
\def\zprime{z}

\shorttitle{SDSS Photometric Covariance}
\shortauthors{Scranton et al.}

\begin{document}

\title{Photometric Covariance in Multi-Band Surveys: Understanding the 
Photometric Error in the SDSS}

\author{Ryan Scranton\altaffilmark{1}, Andrew J. Connolly\altaffilmark{1}, 
  Alexander S. Szalay\altaffilmark{2}, Robert H. Lupton\altaffilmark{3}, 
  David Johnston\altaffilmark{3}, Tam\'as Budav\'ari\altaffilmark{2},
  John Brinkman\altaffilmark{4}, Masataka Fukugita\altaffilmark{5,6}}

\altaffiltext{1}{University of Pittsburgh, Department of Physics and 
  Astronomy, 3941 O'Hara Street, Pittsburgh, PA 15260}
\altaffiltext{2}{Department of Physics and Astronomy, The Johns Hopkins 
  University, 3701 San Martin Drive, Baltimore, MD 21218}
\altaffiltext{3}{Princeton University Observatory, Princeton, NJ 08544}
\altaffiltext{4}{Apache Point Obs., P.O. Box 59, Sunspot, NM 88349-0059}
\altaffiltext{5}{Institute for Advanced Study, School of Natural Sciences, 
  Olden Lane, Princeton, NJ 08540, USA}
\altaffiltext{6}{Institute for Cosmic Ray Research, University of Tokyo,
  Kashiwa 277-8572, Japan}

\email{scranton@bruno.phyast.pitt.edu}

\slugcomment{21 May 2005 Draft}

\begin{abstract}


In this era of precision astrophysics many of our scientific conclusions
rely on a detailed understanding of the uncertainties present within a
data set. Often, however, constraints on the time required to under take
an observation mean that our measures of the variance and covariance
associated with a signal rely simply on a single estimate of the noise
within the data. In this paper we describe a detailed analysis of the
photometric uncertainties present within the Sloan Digital Sky Survey
(SDSS) imaging survey based on repeat observations of approximately 200 square
degrees of the sky. We show that, for the standard SDSS aperture
systems ({\petro}, {\model}, {\psf} and {\cmodel}), the errors generated
by the SDSS photometric pipeline under-estimate the observed scatter in
the individual bands. The degree of disagreement is a strong function of
aperture and magnitude (ranging from 20\% to more than a factor of 2).
We also find that the photometry in the five optical bands can be highly 
correlated for both point sources and galaxies, depending on the aperture and 
magnitude, although the correlation for point sources is almost entirely due 
to variable objects.  Without correcting for this covariance a naive estimate 
of the errors on the SDSS colors could be in error by a factor of two to three.
For the photometric uncertainties on the colors as measured by SDSS photometric
pipeline the strong covariance is cancelled, to some extent, by an 
underestimate of the photometric errors.  As a result, the SDSS errors on the 
colors differ from the observed color variation by approximately 10-20\%
for most apertures and magnitudes.  To facilitate the use of the true 
photometric uncertainties within the SDSS data we provide a prescription 
to correct the errors derived from the SDSS photometric pipeline as a 
function of magnitude (for stars and galaxies) as well as a 
semi-analytic method for generating the appropriate covariance
between the different photometric passbands. Finally, we note that this 
analysis is not specific to just the SDSS photometric survey.  Given the 
strength of the covariance between photometric passbands and the
intrinsic nature of this correlation, we expect that all current and
future multi-band surveys will also observe strongly covariant
magnitudes. Further, since the ability of these surveys to complete
their science goals is largely dependent on color-based target selection
(e.g.\ for selecting QSOs or high redshift galaxies) and photometric
redshifts, these results show the importance of spending a significant
fraction of early survey operations on re-imaging to empirically
determine the photometric covariance of any observing/reduction
pipeline.

\end{abstract}


\section{Introduction}\label{sec:introduction}

Since the earliest astronomical observations using photographic plates,
photometric and color information has been used to characterize and
classify sources (e.g.\ temperatures of stars, stellar populations in
galaxies, identification of QSOs and photometric redshifts of galaxies).
How we interpret these classifications depends on how we account for the
uncertainties present within the photometric measures. For photographic
plate based observations the uncertainties associated with the
photometry (both systematic and statistical) could be substantial.
Consequently, simple estimates of the noise on a measure were often
sufficient to characterize the uncertainties in an analysis.  With the
advent of linear detectors, the shot noise and systematics associated
with the photometry have improved dramatically; as has our ability to
measure magnitudes by including information about an individual object's
morphology or the optical response of the imaging system (i.e.\ the size
and shape of the observed point-spread function). These advances in
photometric precision have led to enhancements in our classification
techniques and in scientific analyses that we can undertake. Large area
surveys such as the Sloan Digital Sky Survey (SDSS) have demostrated the
impact of these improved photometric measures through their use of
multi-band imaging to identify likely QSOs, white dwarfs, luminous red
galaxies for spectroscopic follow-up (cf. \citealt{eis01} and 
\citealt{ric02}).

If we are to fully understand the nature of these selections as well as
to apply other photometric techniques (e.g.\ photometric redshifts) we
must understand not only the errors in each observed band, but also the
errors on the colors generated from combinations of those bands. The
former is calculable given an estimate of the flux from a given object
and the associated sky emission, but the latter will depend on the
relation between the apertures in each of the individual passbands.  In
this paper, we take advantage of the multiple-epoch data available in
the SDSS to measure the photometric covariance matrices for the various
apertures used in the SDSS photometric pipeline as a function of
magnitude, color and object type.  Using these matrices, we examine the 
effect of the covariance between the various bands on the
errors for the typical colors used in object selection.  We compare the
observed scatter in our repeat observations against the scatter expected
from the magnitude errors generated by the photometric pipeline for all
combinations of aperture, magnitude, color, and object type.   The repeat
observations also allow us to extract a sub-population of variable stellar
objects and compare their observed scatter and covariance to the remainder
of the sample.  Finally, we use these comparisions to derive empirical 
relations to correct the photometric and color errors to more accurately 
represent the uncertainties present within the data.

\section{Method}\label{sec:method}

To measure the photometric covariance, we must follow a number of steps.
First, we need to define a clean sample of unique objects with a sufficient
number of well-measured epochs to constrain the full covariance matrix.
Next, we need to convert all of the magnitudes into linear fluxes where 
we can easily compute our statistical means and variances.  We must
define some quantity to measure the relationship between the observed scatter
in the various epochs for a given object to the quoted error that is given
by the SDSS photometric analysis pipeline.  In addition, we need to choose 
criteria for splitting up the sample along a number of different axes 
(brightness, color, object type and variability) to determine which behavior
is universal a which is merely characteristic of a sub-population.  Finally,
all of these analyses need to be repeated for each of the various apertures 
output by the SDSS photometric pipeline.

\subsection{Data}\label{sec:data}

The SDSS photometric system (\citealt{yor00},~\citealt{gun98},~\citealt{smi02},
~\citealt{hog01};~\citealt{ive04}) consists of an array of 30 CCDs arranged in 
six columns ({\it scanlines}) of five CCDs, one for each of the SDSS 
photometric bands ($\uprime$, $\gprime$, $\rprime$, $\iprime$ and $\zprime$;
~\citealt{fuk96}).  During normal operations, the sky passes along the length 
of each scanline giving nearly simultaneous observation in each band.  The 
length of sky in each scanline is further broken up into smaller segments, 
{\it fields}.  A given pass across the sky ({\it strip}) leaves gaps between 
each of the scanlines which are filled in by a strip shifted by the width of 
one scanline.  Combining complementary strips ({\it north} and {\it south}) 
results in a {\it stripe} roughly 2.5 degrees wide.

\begin{figure}
  \begin{center}
    \plotone{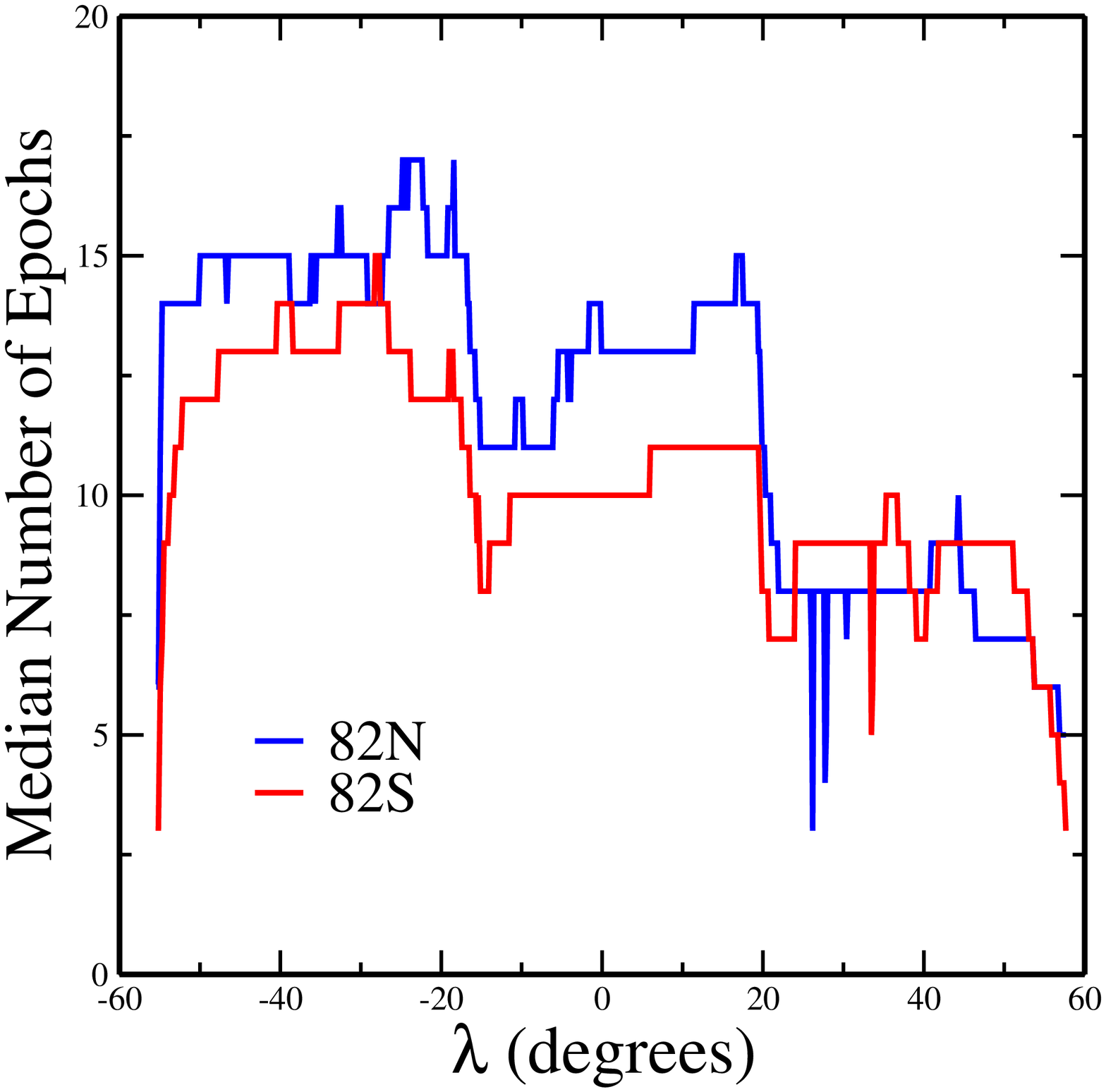}
  \end{center}
  \caption{Median number of epochs for each strip as a function of survey 
    coordinate $\lambda$.}
  \label{fig:epochs}
\end{figure}

As part of normal SDSS operations, the Southern Equatorial stripe has
been scanned repeatedly; each idependent scan is assigned a unique {\it
run} number.  This stripe is centered along 0 $DEC$, running from -51 to
59 degrees in $RA$ (J2000).  In SDSS survey coodinates, the stripe runs
from $-56 < \lambda < 58$.  Figure~\ref{fig:epochs} shows the number of
repeat scans ({i.e.\ \it epochs}) for the central scanlines in the two
strips contained in the Equatorial stripe.  For the purposes of this
project, we restrict ourselves to the areas at least 10 epochs deep.

Within this region, we perform a simple position matching amongst the
objects identified in each run by the photometric pipeline processing
software ({\it photo\_v5.4}; {\photo}, hereafter) using a tolerance of
0\farcs5 for each match.  Since we are not concerned about completeness 
for our sample, we exclude any objects flagged by the pipeline as 
{\tt SATUR}, {\tt SATUR\_CENTER}, {\tt BRIGHT}, {\tt EDGE} {\tt BLENDED} 
and {\tt NOPETRO\_BIG} in any of the five bands.  These cuts reduce the total
number of objects by roughly 10\%.

These flag cuts and are sufficient to eliminate most questionable reductions
and the angular matching tolerance is sufficient for isolated objects 
(\citealt{pie03}).  However, objects whose images are blended together 
represent a special challenge to {\photo}.  The pipeline tags these objects
with the {\tt CHILD} flag and they represent roughly 20\% of the 
catalog remaining after the aforementioned flag cuts.  Since blended objects
are a sizeable fraction of the acceptable objects (particularly for bright
galaxies, where they comprise over half the population) and the effects of 
deblending on the observed scatter are potentially important for the SDSS as 
well as future surveys, we do not exclude them from our main analysis.  
Rather, we will include a parallel analysis focusing solely on isolated 
objects so as to separate the effects of deblending from the remainder of 
the pipeline. 

Finally, all the magnitudes used in each epoch have been dereddened
using the reddening map of \citet{sch98}.  Galactic extinction does not 
affect the intrinsic scatter of an object since all epochs are observed 
through the same line of sight, but it is necessary to correctly make 
the colors cuts we will discuss later on.  This will not be exactly correct
for stars since their light does not pass through the entire Galaxy, but we
will ignore this distinction for the sake of convenience and uniformity.

\subsection{Apertures}\label{sec:mag_systems}

For photometric objects, there are four relevant apertures used for
calculating magnitudes by the photometric pipeline: PSF magnitudes
({\psf}), model magnitudes ({\model}), composite model magnitudes
({\cmodel}), and Petrosian magnitudes ({\petro}).  The details of each
of these apertures can be found in \citet{sto02}, \citet{aba03}, \citet{aba04} 
and \citet{aba05}. For our purposes, a brief description of each will suffice.

{\psf} generates a magnitude from the flux within the local PSF at the 
position of the object.  Rather than a simple Gaussian, {\photo} decomposes
the light profile of bright stars in a given field into 3 Karhunen-Loeve 
modes.  Given a collection of nearby stars, {\photo} then interpolates these
modes to reconstruct the PSF at any given point on the field.  To measure the
{\psf} magnitude, {\photo} fits a Gaussian to the distribution of flux for
a given object and then corrects that flux by applying the same Gaussian
to the reconstructed PSF.  Since it only uses the flux within the PSF, this 
aperture is obviously inappropriate for extended objects, but it does provide 
excellent magnitudes for stellar objects.

For {\model}, PSF-convolved exponential and deVaucouleurs profiles
are fit to the flux distribution in the $\rprime$ band.  The best fitting of
these two profiles is used to calculate the magnitude in each of the five 
bands.  Since the same aperture is used in each band, this is the preferred 
magnitude system for calculating galaxy colors.

{\cmodel} is a variation on {\model}.  As before, exponential and deVaucouleurs
profiles are fit to the flux distribution in the $rprime$ band, making the 
{\exp} and {\dev} magnitudes, respectively.  With these models in place, a 
second fit is performed to find the optimal combination of the two 
models to match the observed flux distribution in each band.  The fractional 
contribution for each aperture in each band is stored in the 
{\tt fracDeV} parameter and the flux for the {\cmodel} ($f_\cmodel$) is given by
\begin{equation}
  f_\cmodel = {\tt fracDeV} f_{\dev} + (1-{\tt fracDeV}) f_{\exp},
\end{equation}
where $0 \le {\tt fracPSF} \le 1$.  Since this prescription attempts to capture
the total flux in each band, it is the preferred aperture for relatively 
faint photometric objects, particularly in the bands other than $\rprime$.  
However, due to the different aperture sizes, it is not appropriate for colors.

{\petro} is a simple flux aperture whose truncation radius is determined by 
finding the radius at which
\begin{equation}
  R_P(r) \equiv \frac{\int^{1.25r}_{0.8r} dr^\prime 2\pi r^\prime I(r^\prime)/
    \left[ \pi (1.25^2 - 0.8^2)r^2 \right]}{\int^r_0 dr^\prime 2\pi r^\prime 
    I(r^\prime)/\left ( \pi r^2 \right)} = 0.2,
\end{equation}
where $I(r)$ is the azimuthally averaged surface brightness. The
{\petro} magnitude is then defined for an elliptical aperture with
semi-major axis twice that of the Petrosian radius. These
magnitudes are appropriate for relatively bright galaxies where the
projected radius can vary strongly and seeing effects are expected to be
minimal. They approximate a total magnitude and are, consequently, used
primarily for the SDSS spectroscopic galaxy samples. At moderately faint
magnitudes ($\rprime>19$), fluctuations in seeing can lead to very
large apertures, extending well beyond most of the flux from the galaxy.
This can lead to rather striking variations in the magnitude for a given
object as well as much larger photometric uncertainties than is seen
when using either {\model} or {\cmodel}.

\subsection{Coaddition} \label{sec:coaddition}

While magnitudes allow for easy numerical descriptions, in order to calculate
the proper covariance between the bands as well as doing a correct coaddition,
we need to convert from {\it asinh} magnitudes (\citealt{lup99}) into flux.  
For a given band $i$, the conversion of magnitude ($m_i$) into flux 
($f_i$) is given by
\begin{equation}
  f_i = 2 F_0 L_i \sinh \left [ -m_i/P - \ln L_i \right ] 
  \label{eq:mag2flux}
\end{equation}
where $F_0 = 3630.78$ Jy, $P = 1.08574$ and 
$L = [1.4 , 0.9 , 1.2 , 1.8 , 7.4 ] \times 10^{-10}$ for the $\uprime$, 
$\gprime$, $\rprime$, $\iprime$, and $\zprime$ bands, respectively.  
Similarly, we can convert magnitude errors ($\Delta m_i$) into 
flux errors ($\Delta f_i$) using
\begin{equation}
  \Delta f_i = 2 \frac{F_0 \Delta m_i}{P} 
  \sqrt{\sinh^2 \left(-m_i/P - \ln L_i \right) + L_i^2}.
  \label{eq:magerr2fluxerr}
\end{equation}

Once in flux units, we can calulate the mean flux ($\bar{f}_i$)  in each
band $i$ for a given object as well as the covariance between 
each of the bands ($C_{f,ij}$):
\begin{equation} \label{eq:covar_def}
  {\rm C}_{f,ij} \equiv \frac{1}{N} \sum^N_{n=1} 
  \left ( f_{n,i} - \bar{f}_i \right )
  \left ( f_{n,j} - \bar{f}_j \right ),
  \label{eq:flux_covar}
\end{equation}
where $f_{n,i}$ is the $n$th epoch measurement of the flux in band $i$ and 
$N$ is the total number of epochs.  When calculating the covariance matrix, it
is important to avoid contamination by random superpositions of unrelated 
objects.  Inadvertantly including a much brighter or fainter object in 
Equation~\ref{eq:flux_covar} will affect the mean flux, as well as leading to 
disproportionately strong off-diagonal elements.  Our tight position matching 
criteria and removal of blended objects is sufficient to exclude the vast
majority of these cases.  To limit the remainder of contamination, we require
that each epoch be within 1 luptitude of the mean. 

In addition to the covariance matrix, we also will need the regression 
matrix ($R_{ij}$):
\begin{equation}
{\rm R}_{ij} \equiv \frac{{\rm C}_{f,ij}}{\sqrt{{\rm C}_{f,ii}{\rm C}_{f,jj}}}
\end{equation}
where we normalize by the variance in each passband.  The regression matrix 
allows us to calculate a number of important quanities.  Most importantly, 
we can convert the diagonal elements of the flux covariance matrix into 
magnitude errors ($\delta m_i$) using the inverse of the transform from 
Equation~\ref{eq:magerr2fluxerr} and use the regression matrix to generate 
the magnitude covariance matrix: 
\begin{equation}
{\rm C}_{m,ij} = {\rm R}_{ij} \delta m_i \delta m_j.  
\label{eq:mag_covar}
\end{equation}
Using ${\rm C}_{m,ij}$, we can calculate proper color errors:
\begin{eqnarray} \label{eq:covar_error}
  \delta (m_i - m_j)^2 & \equiv & \left (\delta m_i\right)^2 + 
  \left (\delta m_j \right )^2 - 2 {\rm C}_{m,ij} \nonumber \\
  \delta (m_i - m_j)^2 & = & \left (\delta m_i\right)^2 + 
  \left (\delta m_j \right )^2 - 2 {\rm R}_{ij} \delta m_i \delta m_j
\end{eqnarray}

In order to ensure that the variation we observe between each epoch is
unaffected by any possible calibration differences, we must tie the
magnitude zero-points in each run together.  This requires us to choose
one run, the {\it zero run}, from each strip which extends the full
length of the stripe to use as the basis for the zero-points on that
strip (runs 3384 and 4203 for the north and south strips, respectively).
This eliminates the use of these runs for determining ${\rm C}_{f,ij}$ but
prevents the calculation from being dominated by constant off-sets
ranging across all five bands.

To tie the photometric zero-points together, we first divide each scanline
into 20 segments.  Within each segment, we find all of the stars with 
{\psf} errors less than 0.05; this typically restricts us to objects
brighter than 17th magnitude in a given band (although considerably deeper 
in $\gprime$, $\rprime$ and $\iprime$).  The stars for a given run
are matched by position against the zero run in the same strip.  We then 
calculate the mean difference in {\psf} for all of the run's stars 
relative to those in the zero run in each band.  These mean differences are 
used as the zero-point offset at the mid-point of that segment.  For 
individual objects within a run, we interpolate based on those mid-points and 
subtract the resulting zero-point from all of the objects in the run.   
Typical values for the zero-point offset range from -0.02 to 0.02 magnitudes.  

\subsection{Selection Cuts}\label{sec:selection_cuts}

For the purposes of dividing our data set, we convert $\bar{f}_i$ into a
coadded magnitude ($\bar{m}_i$) for each unique object.  The most basic cut 
is a simple magnitude selection, dividing the sample into unit magnitude 
slices from 17 to 21 in $\rprime$.  Within each magnitude slice, we also 
separate the sources into red and blue objects.  A number of possible cuts 
have been used in previous papers for this purpose.  In this analysis, we 
use the cut determined by \citet{bal04}:
\begin{eqnarray}
  \bar{u} - \bar{r} &<& 1.8 ~~:~~  {\rm red} \\ \nonumber
  \bar{u} - \bar{r} &>& 1.8 ~~:~~  {\rm blue}
  \label{eq:color_sep}
\end{eqnarray}
Finally, we divide objects into stars and galaxies.  In the coadded
magnitudes, there is a very clean separation of galaxy and stellar loci
using the concentration parameter, $c = \bar{m}_\psf - \bar{m}_\cmodel$.
A simple cut at
\begin{eqnarray}
  c &<& 0.05 ~~:~~ {\rm star} \\ \nonumber
  c &>& 0.1 ~~:~~ {\rm galaxy}
\end{eqnarray}
in $\rprime$ will divide the sample without significant contamination in
either group beyond our faint limit at $\rprime = 21$.  Combining all of 
these cuts, we have 24 total sub-samples to consider in each band and each 
aperture.  See Table~\ref{tab:obj_counts} for a listing of the number of 
unique objects in each sub-sample for each aperture.

\subsection{Comparison to Pipeline Errors}\label{sec:pipeline}

Properly computed photometric errors should reflect the variation in
subsequent measurements of the same object.  To compare the
epoch-to-epoch scatter to the magnitude errors coming out of the
photometric pipeline, we calculate the following quantity:
\begin{equation} \label{eq:chi2}
  \chi^2_{F,i} = \sum^{N}_{n=1} 
  \left ( \frac{f_{n,i} - \bar{f}_i}{\Delta f_{n,i}} \right )^2 ,
\end{equation}
where we have used the nomenclature from Equation~\ref{eq:covar_def} and
$\Delta f_{n,i}$ is the photometric pipeline flux error on the $n$th
epoch observation in band $i$.  For correctly calculated errors,
$\chi^2_F$ should be a $\chi^2$ distribution peaked around the number of
degrees of freedom ($N$).  An equivalent quantity for the magnitudes,
($\chi^2_M$), can be calculated using the epochal magnitudes and errors
as well.  To the extent that the transformations in 
Equations~\ref{eq:mag2flux} and~\ref{eq:magerr2fluxerr} are valid, 
$\chi^2_F$ and $\chi^2_M$ should be equivalent.

To characterize how well the pipeline errors describe the true
uncertainties for all of objects within a given selection bin, we sum
the values of $\chi^2_{M,i}$ for all of the $N^\prime$ objects within
that bin and compare it to the sum of the degrees of freedom:
\begin{equation}
\chi^2_{T,i} = \Sigma^{N^\prime}_{n=1} \left(\chi^2_{M,i}\right)_n~~;~~
N_T = \Sigma^{N^\prime}_{n=1} \left( N \right)_n,
\end{equation}
where $(\chi^2_{M,i})_n$ is the magnitude form of Equation~\ref{eq:chi2} for
object $n$.  Since $\chi^2_{T,i}$ is also a $\chi^2$ distribution, we expect 
that
\begin{equation}
\chi^2_{N,i} \equiv \chi^2_{T,i}/N_T
\end{equation}
should be near unity for well-characterized photometric pipeline errors.
Further, the mean ratio between the pipeline errors and the actual observed 
scatter should be given by ${\cal R}_i \equiv\sqrt{\chi^2_{N,i}}$ in each bin.

\subsection{Variable Objects}\label{sec:variability}

While the methology described above is appropriate for objects where the 
scatter in successive measurements is due to statistical fluctuations, 
intrinsic variability can affect our results in two ways.  First, this will 
bias our comparison of the observed scatter to the errors, increasing the 
disagreement between the two.  Second, since the variability in these objects
typically happens over a broad spectral range, this will create strong 
correlations between filters which would otherwise be independent (or at least
more weakly correlated).  Including even a relatively small fraction of 
strongly variable objects in a given magnitude/color/object type bin with 
otherwise uncorrelated filters (as we would expect for stellar objects, for 
example) can result in moderately strong off-diagonal regression matrix 
elements for the ensemble average.

To remove these objects, we matched our 10+ epoch objects against the 
catalog of photometric quasars described in \citet{ric04}.  Quasars show
strong photometric variability (\citealt{van04}) and the corresponding 
regression matrices for these objects have very large off-diagonal elements
(see \S\ref{sec:regression}).  This makes the median regression matrix 
determinant for these objects very small ($\sim 0.003$), compared to that of 
{\model} galaxies ($\sim 0.01$) and the total stellar population ($\sim 0.2$).
By selecting objects with regression matrix determinant less than 0.008, 
we can effectively split our stellar population into variable 
($\sim$ 5-10\% of the total) and non-variable sub-populations.  A similar
cut for galaxies using {\model} selects roughly half the sample and the 
``non-variable'' galaxies still show significant off-diagonal elements.  Based
on this and the fact that we expect little intrinsic variation in galaxy 
spectral distribution on the time scale of our repeat observations 
(relatively speaking), we will only consider variability for stellar objects.

\section{Results}\label{sec:results}

\subsection{Pipeline Errors vs. Observed Scatter}\label{sec:chi2_result}

\begin{figure}
  \begin{center}
    \includegraphics[height=6.5in]{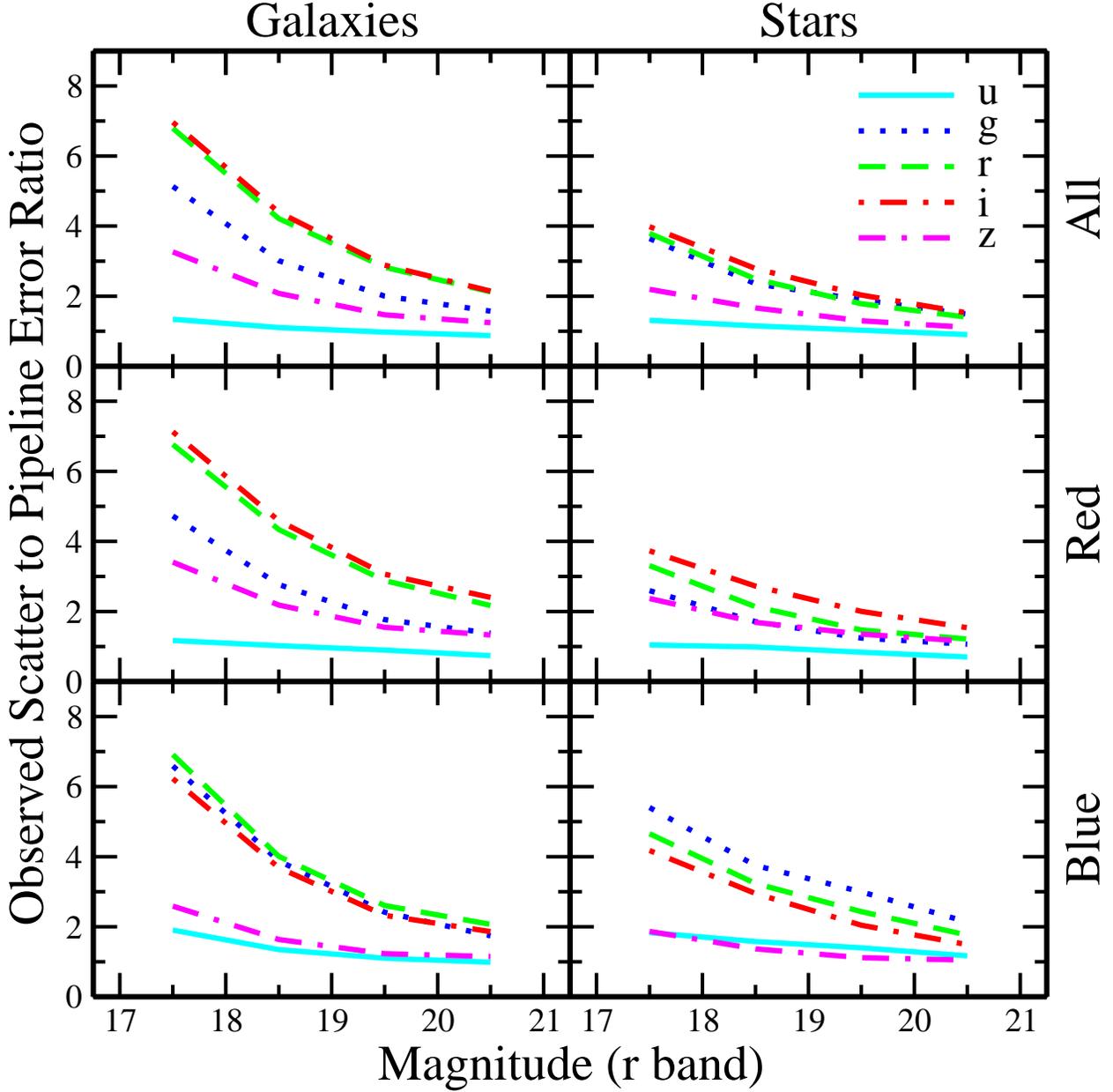}
  \end{center}
  \caption{Ratio of observed magnitude errors to {\photo} pipeline errors 
    ($\chi_{\rm N}$) as a function of object type, magnitude and color using 
    {\model}. The $\uprime$ ratio is indicated by the solid lines, $\gprime$ 
    by the dotted line, $\rprime$ by the dashed line, $\iprime$ by the
    dot-dashed line, and $\zprime$ by the dash-dotted line.}
  \label{fig:mag_error_model}
\end{figure}

\begin{figure}
  \begin{center}
    \includegraphics[height=6.5in]{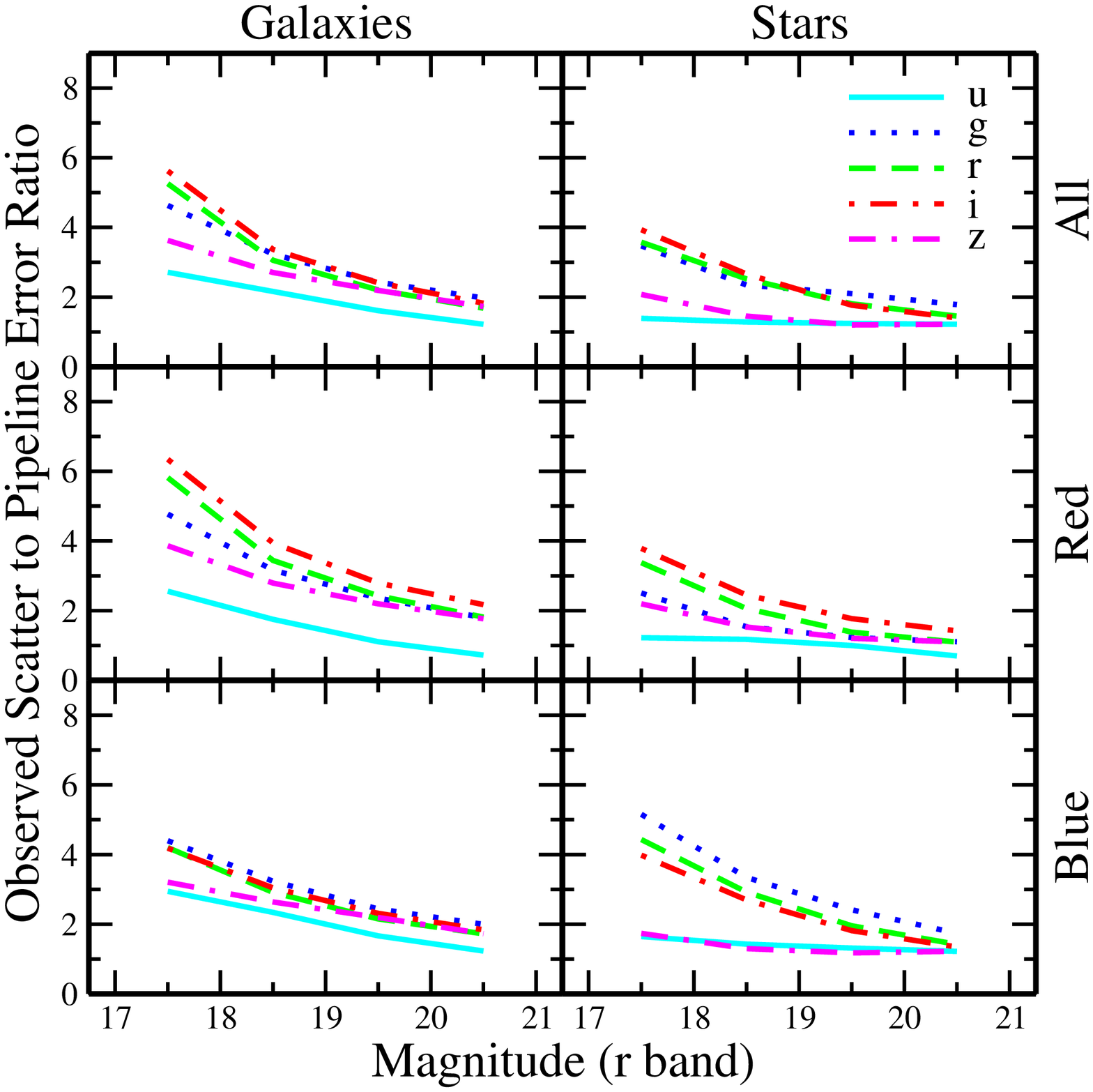}
  \end{center}
  \caption{Same as Figure~\ref{fig:mag_error_model}, but using {\cmodel}.}
  \label{fig:mag_error_cmodel}
\end{figure}

\begin{figure}
  \begin{center}
    \includegraphics[height=6.5in]{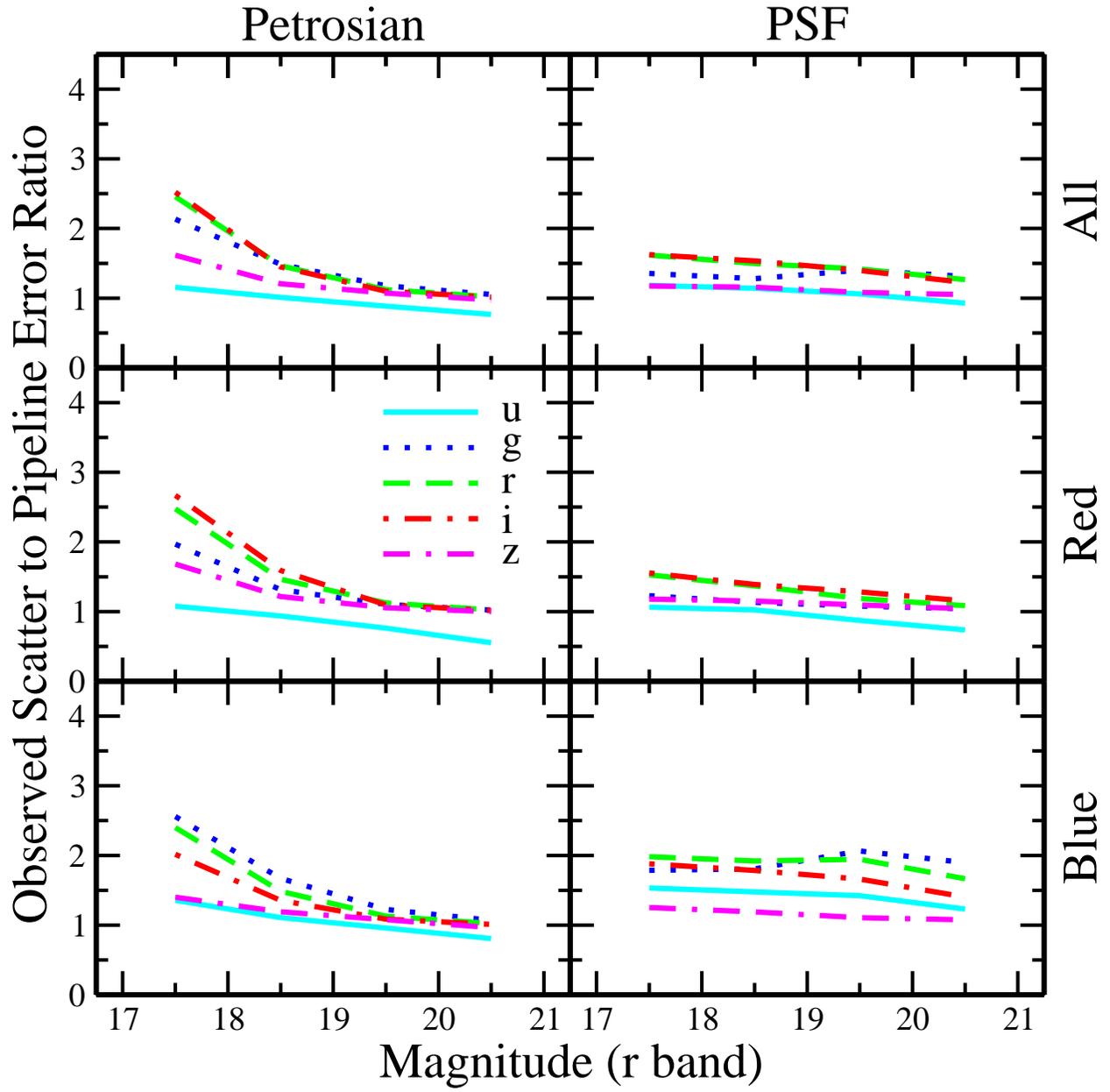}
  \end{center}
  \caption{Same as Figure~\ref{fig:mag_error_model}, but for galaxies
    using {\petro} and stars using {\psf}.}
  \label{fig:mag_error_petro_psf}
\end{figure}

Tables~\ref{tab:chi2_model} through~\ref{tab:chi2_psf} show the values
of $\chi^2_{\rm N}$ for the various selection criteria and
Figures~\ref{fig:mag_error_model} through
\ref{fig:mag_error_petro_psf} show ${\cal R}$, the ratio between the
observed scatter and the {\photo} pipeline errors, for each band.  In
general, we find that bright objects typically display much larger
scatter than one would expect based on the {\photo} errors; the values of 
$\chi^2_{\rm N}$ in Tables~\ref{tab:chi2_model} through~\ref{tab:chi2_psf} are
consistent with the peaks in the various distributions of $\chi^2_F$, 
confirming our contention in \S\ref{sec:pipeline} that the aggregate 
would be $\chi^2$ distribution.  The excess scatter was most prominent with 
{\model} and least with {\psf}, but exists for all apertures.  It should be
noted, however, that despite the seemingly large ratios between the observed
scatter and the pipeline errors, the observed scatter at the bright end remains
very small in an absolute sense (this can be easily inferred from the values 
of the color errors given in Tables~\ref{tab:color_error_model_gal} through~
\ref{tab:color_error_psf_star} described in \S\ref{sec:color_errors}).  As 
objects become fainter, the agreement between {\photo} errors and the 
observed scatter generally improved.  The $\uprime$ band scatter also tended 
to be much better matched by the pipeline errors at all magnitudes in all 
apertures (see \citealt{bal05} for more details on $\uprime$ band errors).  
This behavior is consistent with the notion that the bright end scatter can be 
strongly influenced by the details of modelling the light distribution, 
while at the faint end photon noise becomes the dominant source of scatter.  
Finally, in all apertures, stars showed a much stronger color dependence on 
$\chi^2_{\rm N}$ than galaxies, with blue stars usually giving a much larger 
observed scatter relative to their {\photo} errors than red stars (see 
\S\ref{sec:variability_results} for more details).

For {\model}, the values of $\chi^2_{\rm N}$ tended to be most extreme
in the $\rprime$ and $\iprime$ bands.  In general, the {\photo} errors
were under-estimated by at least a factor of 2 for almost all
combinations of magnitude, object type and color, and by as much as a
factor of 6 in some cases.  Stars, however, typically had smaller values
of $\chi^2_{\rm N}$ for a given magnitude and color.  Like {\model}, the
{\photo} errors for {\cmodel} were, in general, strongly under-estimated
relative to the observed scatter, albeit to a lesser degree than
{\model}.  {\cmodel} also showed the same $\chi^2_{\rm N}$ split between
stars and galaxies.  {\petro} and {\psf} errors typically showed much
better agreement between the {\photo} errors and the observed scatter
than either of {\model} or {\cmodel}.

\subsection{Regression Matrices}\label{sec:regression}

\begin{figure}
  \begin{center}
    \includegraphics[height=6.5in]{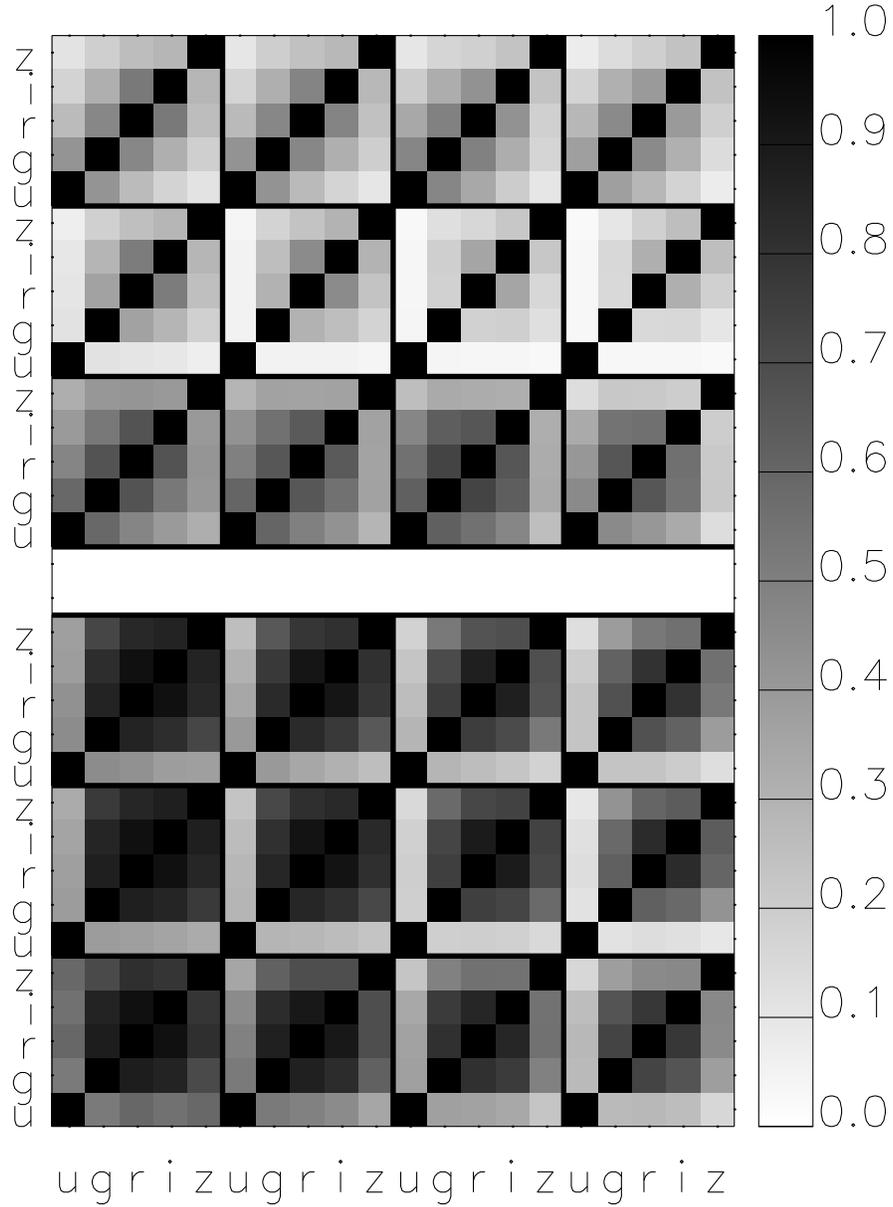}
  \end{center}
  \caption{Regression matrices for objects using {\model}, galaxies in the 
    lower block and stars in the upper block.  The $5 \times 5$ matrices 
    separate the total sample of unique objects into 4 magnitude bins 
    ($17 < \rprime < 18$, $18 < \rprime < 19$, $19 < \rprime < 20$, 
    $20 < \rprime < 21$) and 3 color bins: no color cut, red objects 
    ($\uprime - \rprime > 1.8$) and blue objects ($\uprime - \rprime < 1.8$).
    In each block, the top row gives the regression matrices for objects 
    with no color cut, the middle row for the red objects and the bottom 
    row for the blue objects.  Going from left to right runs from the 
    brightest to faintest magnitude cut in each row.}
  \label{fig:reg_model}
\end{figure}

\begin{figure}
  \begin{center}
    \includegraphics[height=6.5in]{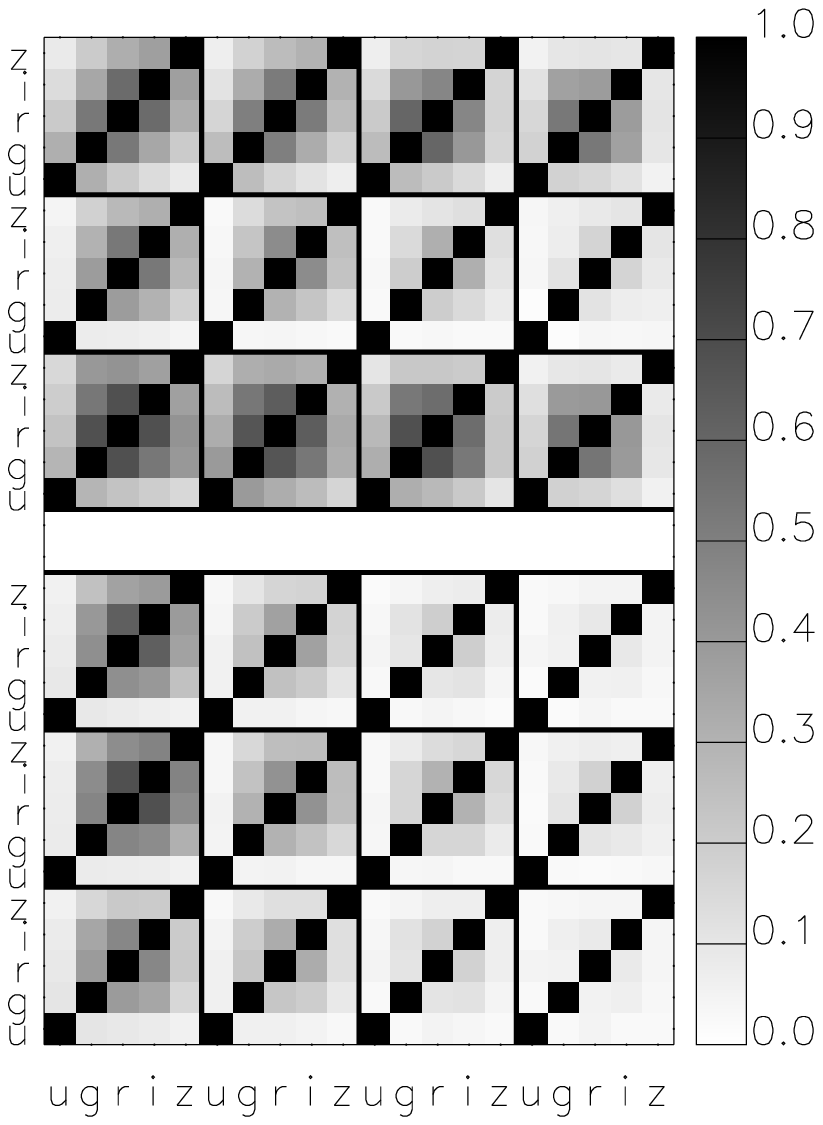}
  \end{center}
  \caption{Same as Figure~\ref{fig:reg_model}, but for {\cmodel}.}
  \label{fig:reg_cmodel}
\end{figure}

\begin{figure}
  \begin{center}
    \includegraphics[height=6.5in]{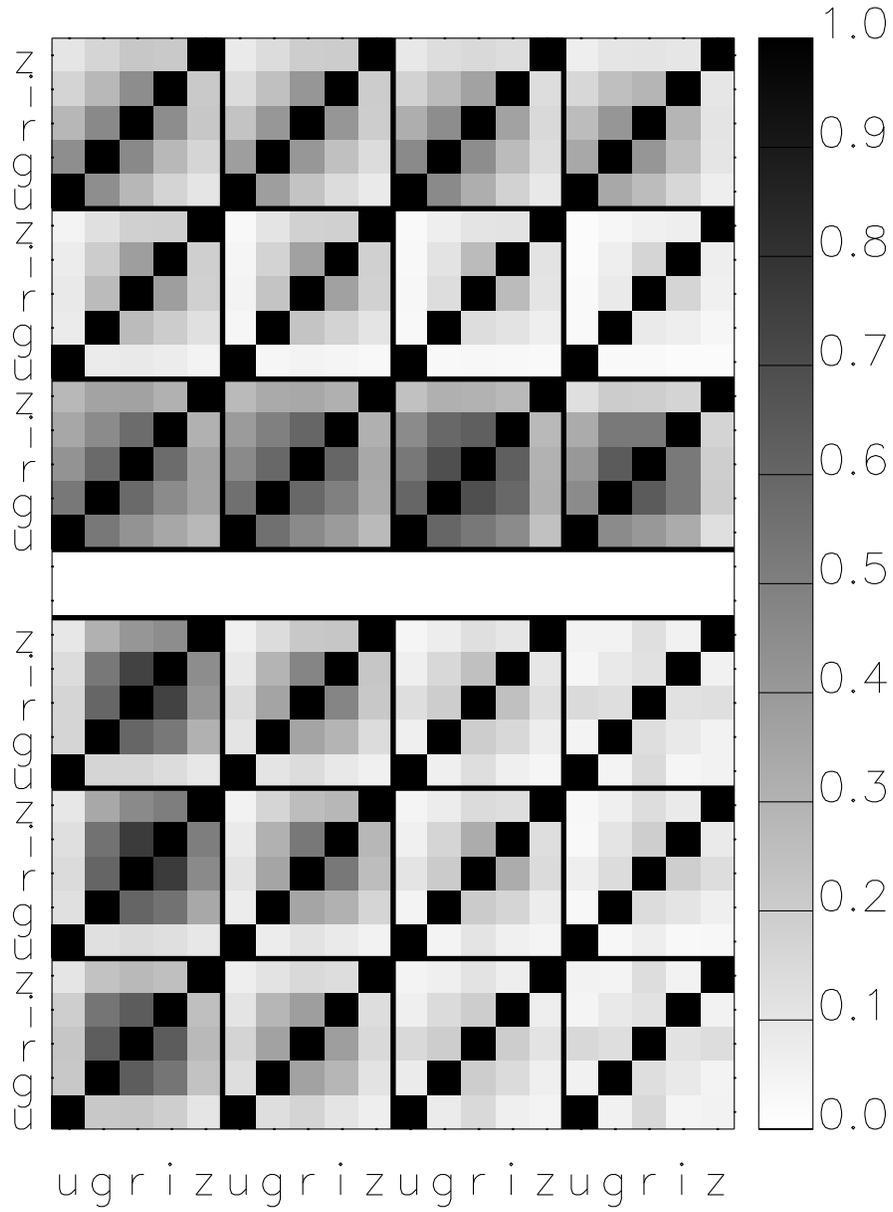}
  \end{center}
  \caption{Same as Figure~\ref{fig:reg_model}, but for {\psf} in the upper
    block and {\petro} in the lower block.}
  \label{fig:reg_petro_psf}
\end{figure}

Figures~\ref{fig:reg_model} through~\ref{fig:reg_petro_psf} show the
mean regression matrices for objects that fall within each selection bin
for each aperture.  We do not plot the regression matrices for galaxies
using the {\psf} magnitudes nor for stars with the {\petro} magnitudes,
as these systems are intrinsically inappropriate for those objects.

The primary determining factor for the strength of the off-diagonal
elements of the regression matrix was object type.  Stars observed using
{\model}, {\cmodel}, and {\psf} produced $R$ matrices that were nearly
identical, with strong off-diagonal elements for blue stars 
($R_{\uprime,\gprime}, R_{\gprime,\rprime}, R_{\rprime,\iprime} \sim 0.4-0.6$) 
and weaker equivalent elements (0.2-0.4) for red stars.  The amplitude
of the off-diagonal elements was somewhat smaller for the {\psf} and
{\cmodel} regression matrices compared to the {\model} regression
matrices, but the relative strength of the elements within each
associated matrix was very similar.  Likewise, all three apertures
showed the same slight evolution toward weaker off-diagonal elements for
fainter samples relative to bright ones.

For extended objects, the strongest off-diagonal elements were found
using the {\model} magnitudes.  This is not surprising given that the
aperture used in this method is not independently fit in each band.
With the exception of $R_{\uprime,\gprime}$, the nearest off-diagonal
elements using {\model} typically ranged from 0.6 to 0.9, with similarly
strong elements in the $R_{\gprime,\iprime}$ elements.  By comparison,
the same terms in the regression matrices from {\cmodel} and {\petro}
were almost always less than 0.5, usually much less.  As with stars, the
off-diagonal amplitude for all apertures decreases as a function of
increasing magnitude.  For {\cmodel} and {\petro}, the evolution is
quite strong relative to that seen in {\model}; at the faintest
magnitudes the regression matrices for {\cmodel} and {\petro} are nearly
diagonal, while the off-diagonal elements for {\model} have diminished
by only 10-15\%.

Separating galaxies by color, we find that the regression matrices for
blue galaxies typically have stronger terms along the $\uprime$ column
than those for red galaxies; clearly these terms are suppressed in the
latter matrices by the relatively faint $\uprime$ magnitudes for those
sources.  The $R_{\gprime,\rprime}$, $R_{\rprime,\iprime}$ and
$R_{\gprime,\iprime}$ terms are generally very similar regardless of
aperture, while red galaxies have stronger $R_{\iprime,\zprime}$ terms.
These tendencies are consistent regardless of magnitude or aperture.

\subsection{Color Errors}\label{sec:color_errors}

To estimate the relative importance of covariance on color errors and to
compare with the {\photo} errors and the observed scatter, we calculate
four quantities for the most commonly used colors ($\uprime - \gprime$,
$\gprime - \rprime$, $\rprime - \iprime$, and $\iprime - \zprime$): \\
\\ 
$\bullet$ {\bf Observed Errors}: The measured error based on the
mulit-epoch data measured independently for each
color for objects in a given magnitude/object type/color
bin. \\ 
$\bullet$ {\bf Proper Errors}: Using the observed scatter in
each band ($\delta m_i$) and the appropriate regression matrix, we
calculate the color error as given in Equation~\ref{eq:covar_error}.  If
the errors are Gaussian, then the proper errors should match the
observed errors closely. \\ 
$\bullet$ {\bf Naive Errors}: Like proper errors, except that the
covariance term is omitted.  The ratio between the proper and naive
errors indicates the strength of the covariance between bands. \\
$\bullet$ {\bf {\photo} Errors}: Using $\chi_N$, we transform the
observed scatter in each band into an estimate of the mean {\photo}
errors for the objects in each bin.  Since we do not have a proper
covariance matrix for the {\photo} errors, we calculate the color errors
without it.

\begin{figure}
  \begin{center}
    \includegraphics[height=6.5in]{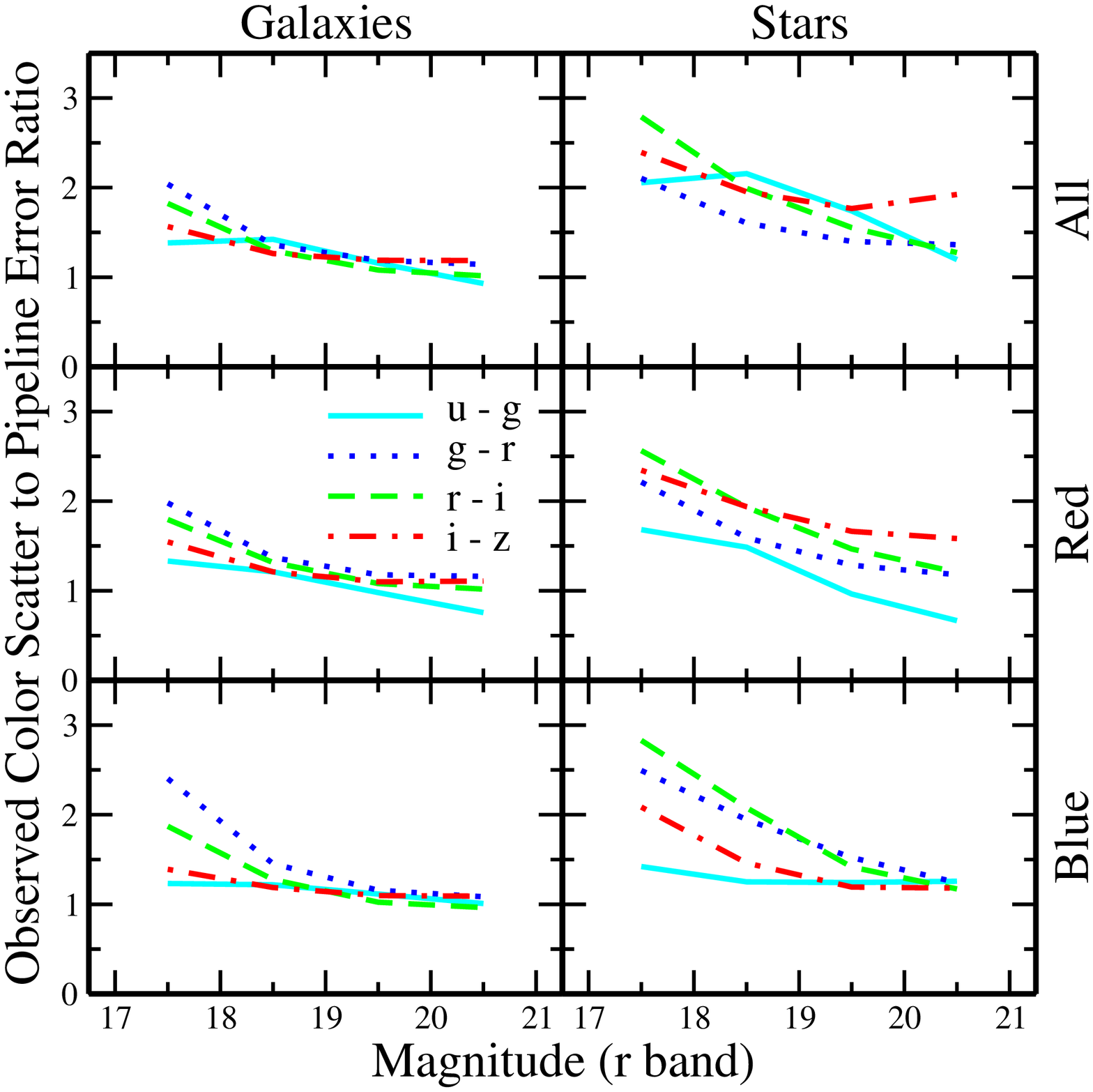}
  \end{center}
  \caption{Ratio of observed color errors to {\photo} pipeline color errors 
    as a function of object type, magnitude and color using {\model}. 
    $\uprime - \gprime$ error ratio is indicated by the solid lines, 
    $\gprime - \rprime$ by the dotted line, $\rprime - \iprime$ by the dashed 
    line, and $\iprime - \zprime$ by the dot-dashed line. }
  \label{fig:color_error_model}
\end{figure}

\begin{figure}
  \begin{center}
    \includegraphics[height=6.5in]{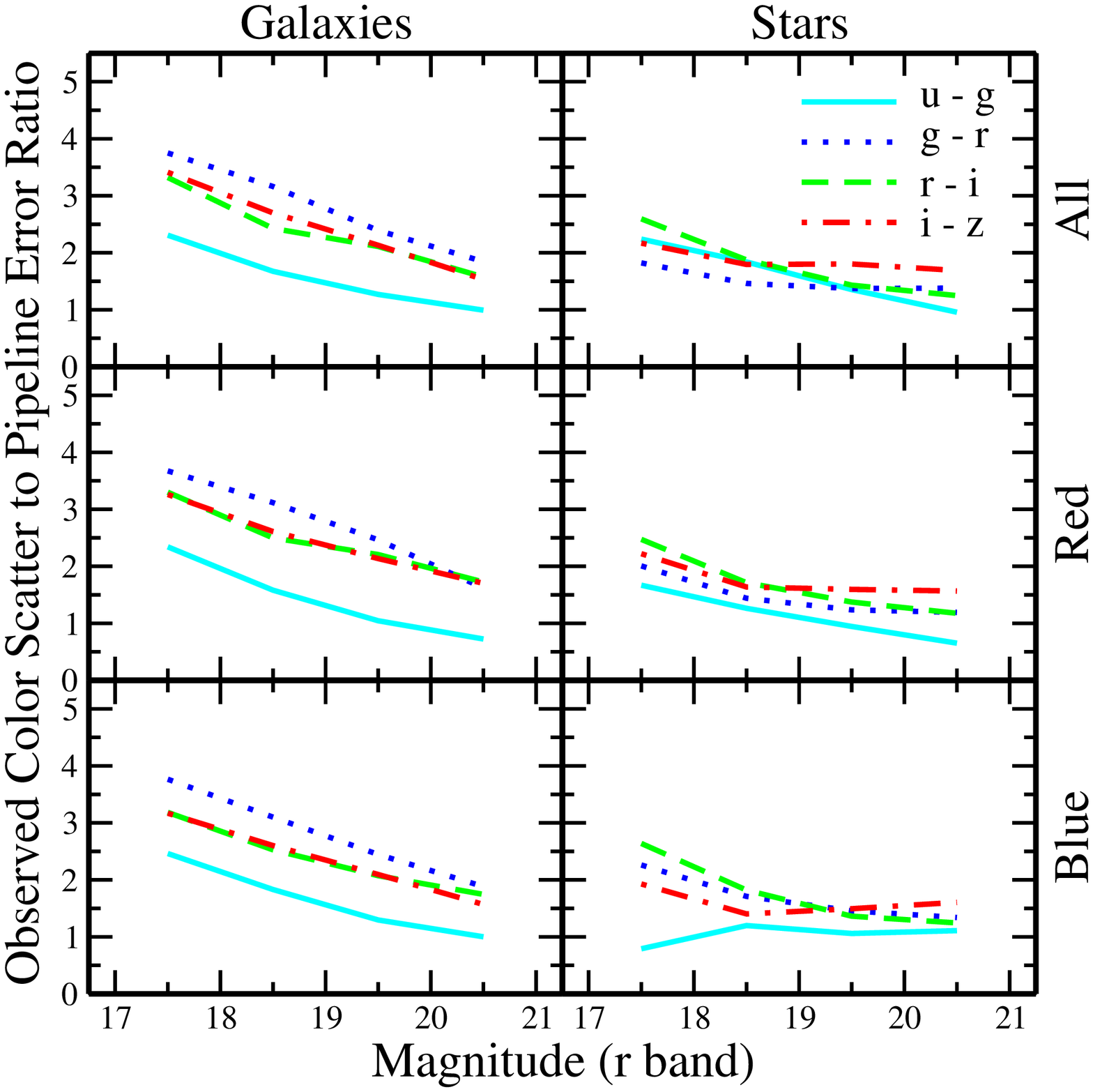}
  \end{center}
  \caption{Same as Figure~\ref{fig:color_error_model}, but using {\cmodel}.}
  \label{fig:color_error_cmodel}
\end{figure}

\begin{figure}
  \begin{center}
    \includegraphics[height=6.5in]{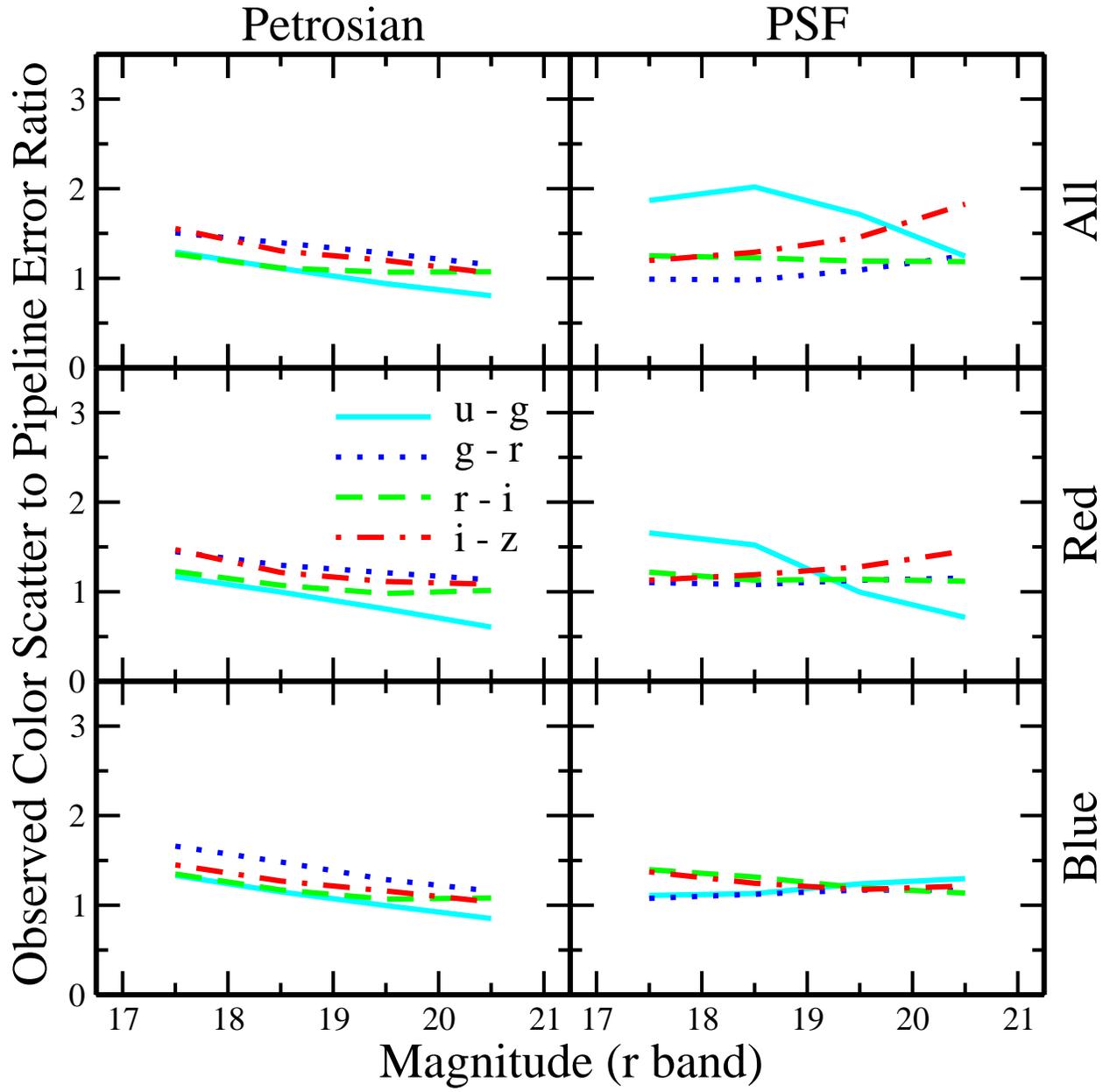}
  \end{center}
  \caption{Same as Figure~\ref{fig:color_error_model}, but for galaxies
    using {\petro} and stars using {\psf}.}
  \label{fig:color_error_petro_psf}
\end{figure}

Tables~\ref{tab:color_error_model_gal} through~
\ref{tab:color_error_psf_star} provide these quantities as a function of
aperture, object type, magnitude, and color.
Figures~\ref{fig:color_error_model}
through~\ref{fig:color_error_petro_psf} show the ratio between the
observed color errors and the {\photo} color errors.  It is worth noting
at the outset that colors measured with apertures other than {\model}
and {\psf} (and only for stars in the latter case) are not meant to be
meaningful, so disagreements between the observed scatter and the
{\photo} color errors are not likely to be relevant to any current or
future research.  For completeness, however, we will touch on them
briefly.

As expected by the strong off-diagonal elements in the galaxy {\model}
regression matrices, the naive errors here were typically larger than
the proper errors by 20-60\%.  With the exception of the
$\uprime-\gprime$ color, the proper errors generally matched the
observed color scatter very well.  Indeed, the match between proper
errors and the observed scatter for the other three colors was very good
regardless of aperture, object type, magnitude or color.  This verifies
that, fundamentally, the scatter in the colors is well modeled by a
Gaussian.  $\uprime-\gprime$ errors were typically under-estimated by
the proper errors, although part of this discrepancy may have been due
to $\uprime$ drop-outs.  Despite the strong covariance between bands
seen in the regression matrices, {\photo} errors matched the observed
scatter nearly as well as the proper errors
(Figure~\ref{fig:color_error_model}), although less so at bright
magnitudes where they tend to under-estimate the color errors.

For {\psf} , the agreement between the proper error, the {\photo} error
and the observed scatter was also quite good.  The {\photo} errors were
typically smallest of the three, but usually by no more than 10\%.
{\photo} errors were particularly good for blue stellar objects
(Figure~\ref{fig:color_error_petro_psf}).

While the difference between {\photo} errors and the observed scatter in
each band generally cancelled the lack of a covariance term for {\model}
galaxies and {\psf} stars, the {\photo} color errors for stars observed
with {\model} were under-estimated by as much as 100\%
(Figure~\ref{fig:color_error_model}).  This was true regardless of
magnitude or object color, although fainter stars tended to have smaller
differences between the {\photo} color errors and the observed scatter.

Likewise, the color errors from {\photo} were under-estimated for all
observed with {\cmodel} magnitudes by factors as large as 2-3.  In all
cases, the color errors were under-estimated by at least 25\%
(Figure~\ref{fig:color_error_cmodel}).  {\photo} color errors were
generally under-estimated by 10-20\% for {\petro}
(Figure~\ref{fig:color_error_petro_psf}).

\subsection{Variability Results}\label{sec:variability_results}

\begin{figure}
  \begin{center}
    \includegraphics[height=6.5in]{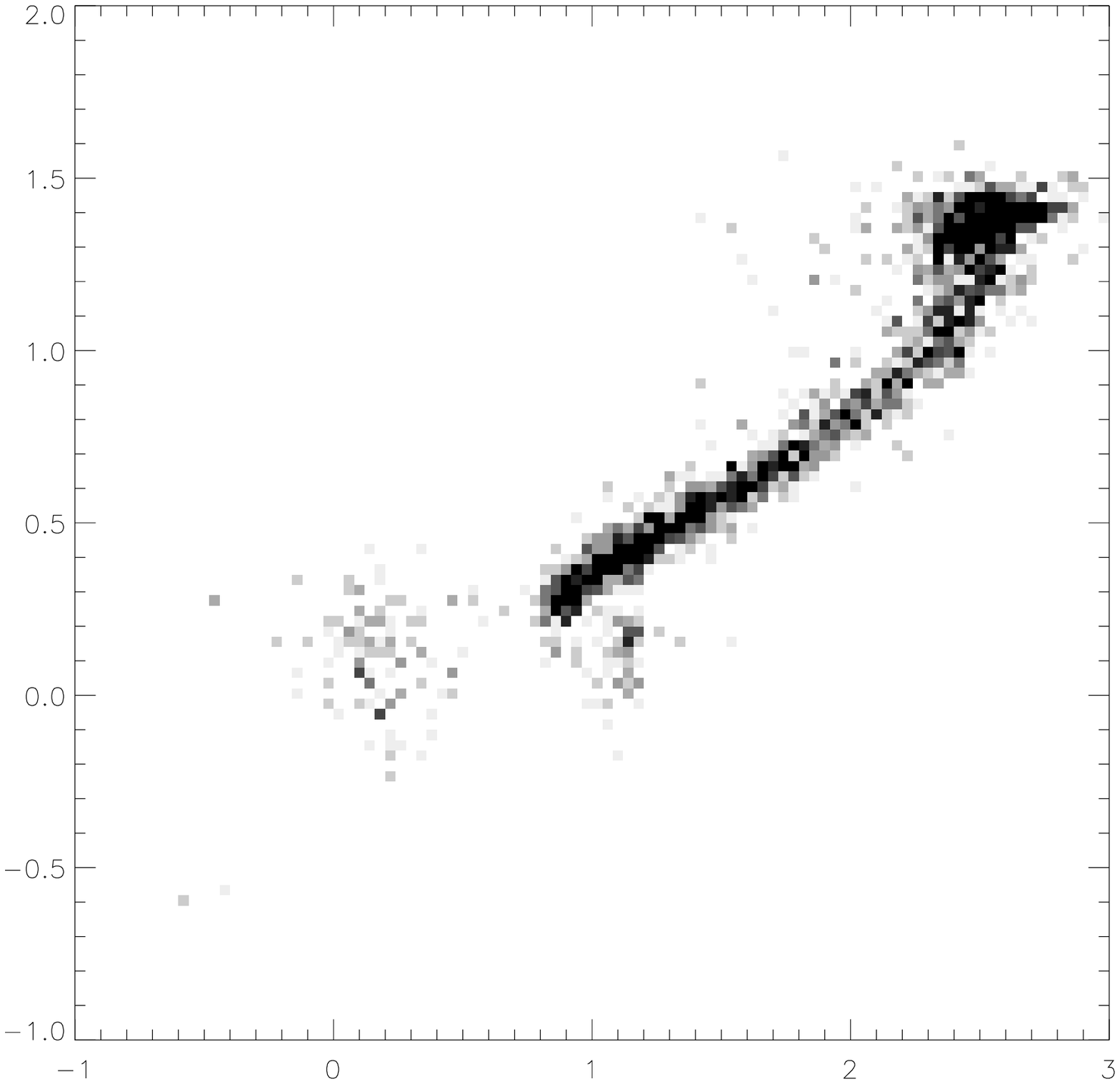}
  \end{center}
  \caption{Coadded $\uprime - \gprime$ vs. $\gprime - \rprime$ for variable
    stellar objects.}
  \label{fig:vary_color_color}
\end{figure}

\begin{figure}
  \begin{center}
    \includegraphics[height=6.5in]{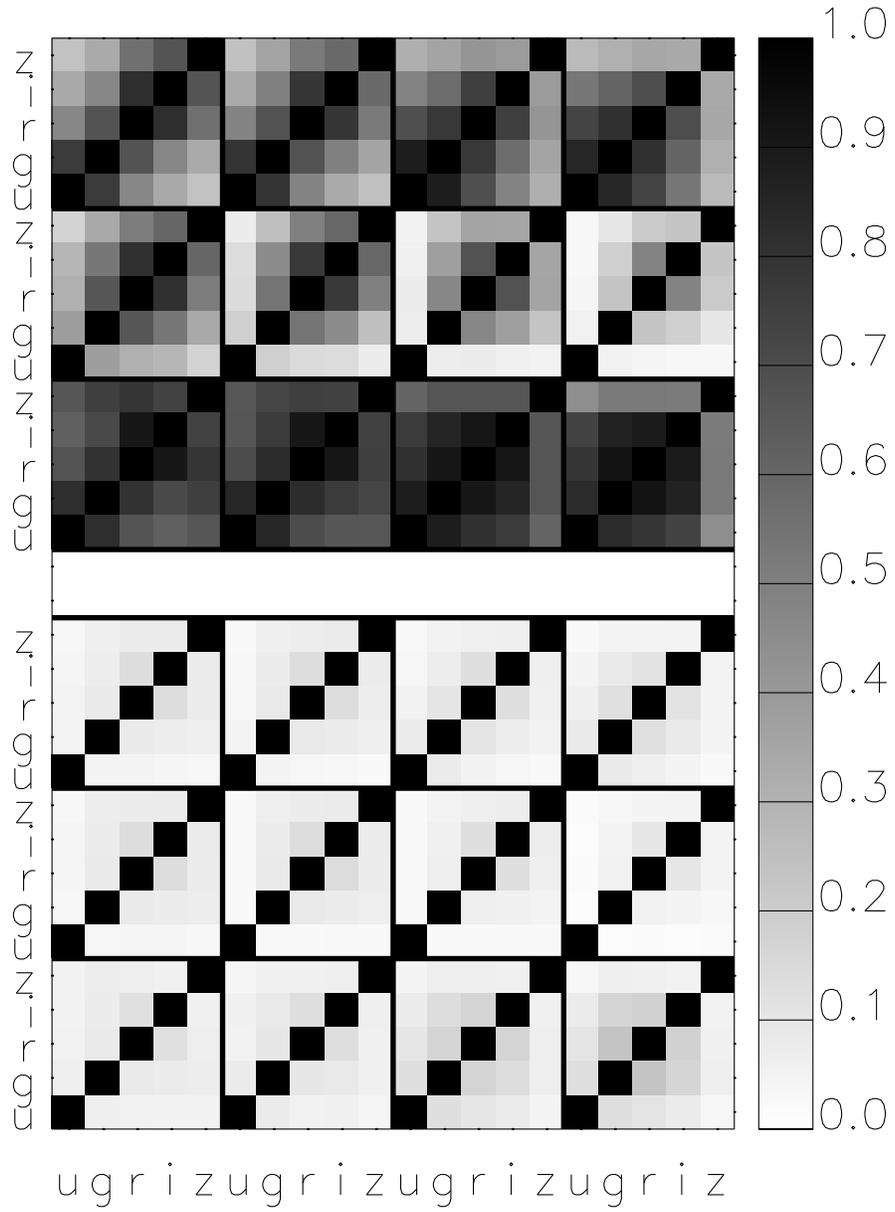}
  \end{center}
  \caption{Same as Figure~\ref{fig:reg_model}, but for variable stars 
    objects using {\psf} in the upper block and non-variable stars in 
    the lower block.}
  \label{fig:reg_psf_vary}
\end{figure}

\begin{figure}
  \begin{center}
    \includegraphics[height=6.5in]{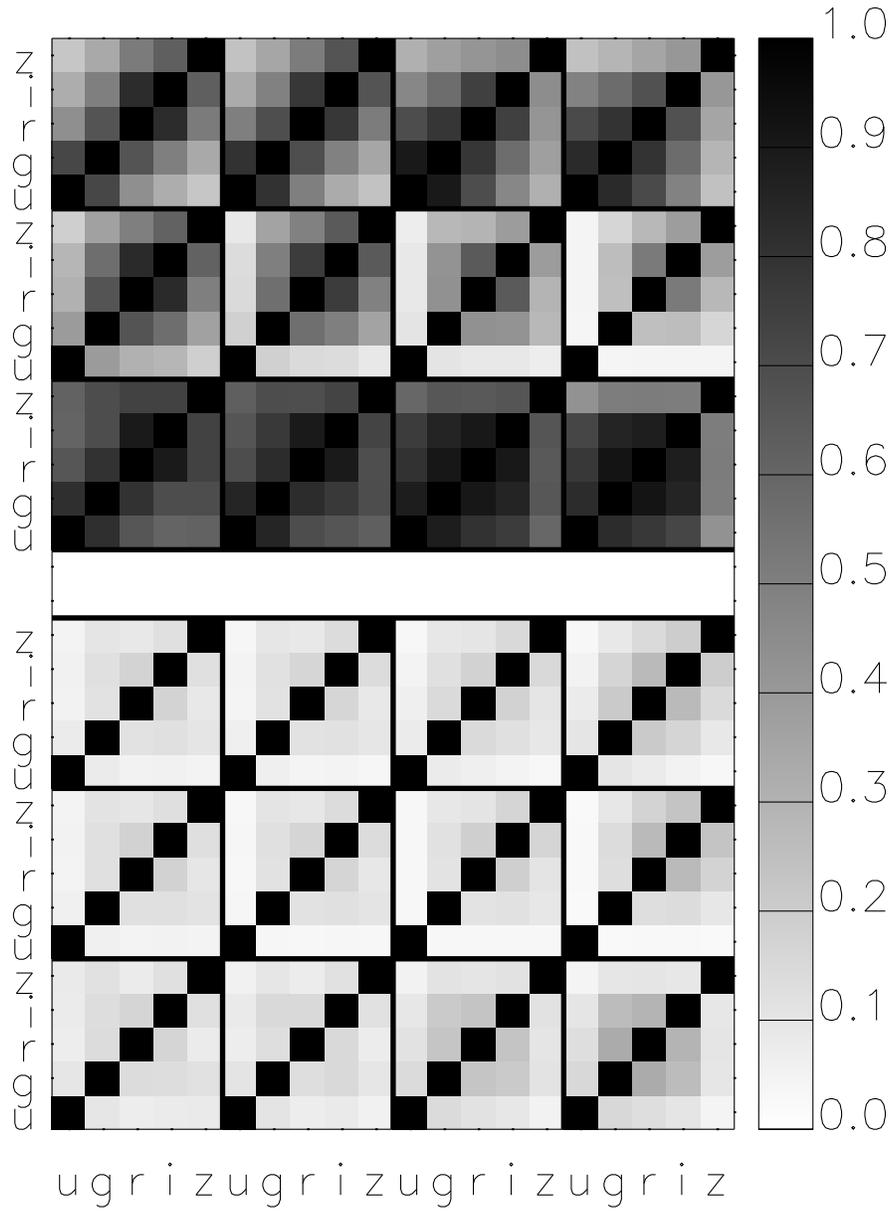}
  \end{center}
  \caption{Same as Figure~\ref{fig:reg_psf_vary}, but using {\model}
    but for variable stars.} 
  \label{fig:reg_model_vary}
\end{figure}

\begin{figure}
  \begin{center}
    \includegraphics[height=6.5in]{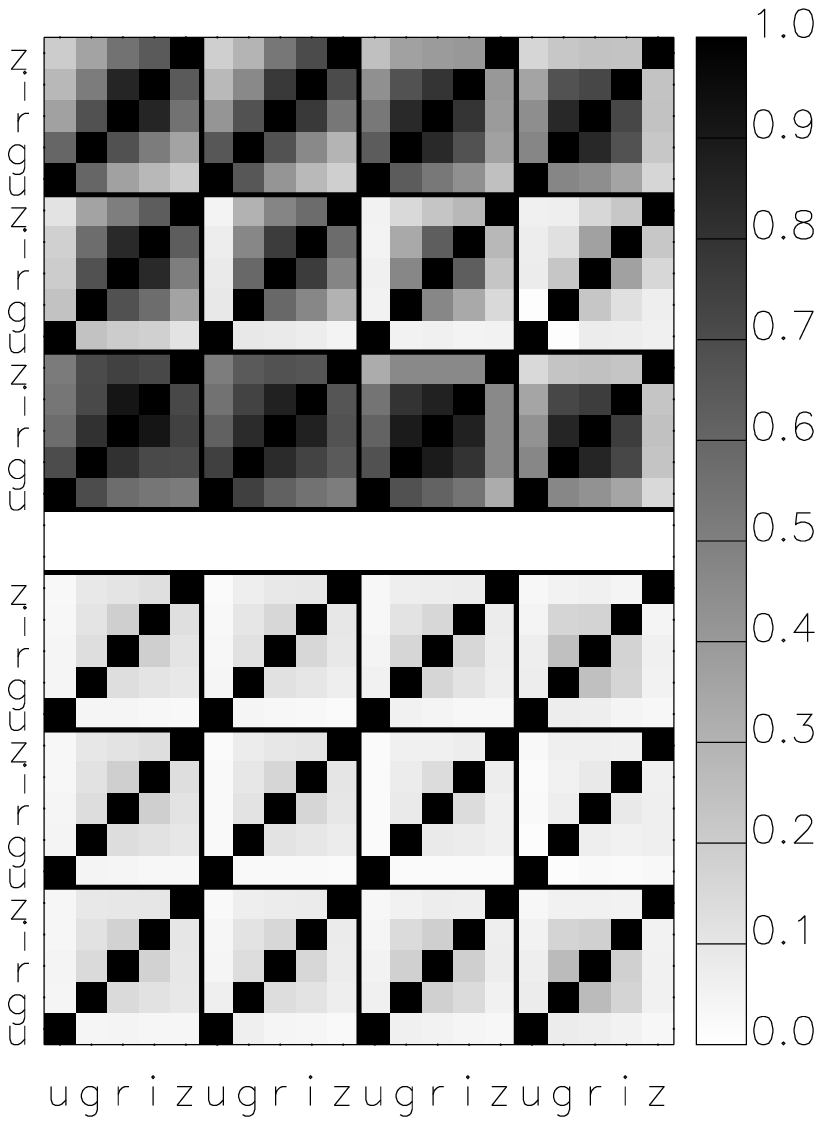}
  \end{center}
  \caption{Same as Figure~\ref{fig:reg_psf_vary}, but using {\cmodel}
    but for variable stars.} 
  \label{fig:reg_cmodel_vary}
\end{figure}

\begin{figure}
  \begin{center}
    \includegraphics[height=6.5in]{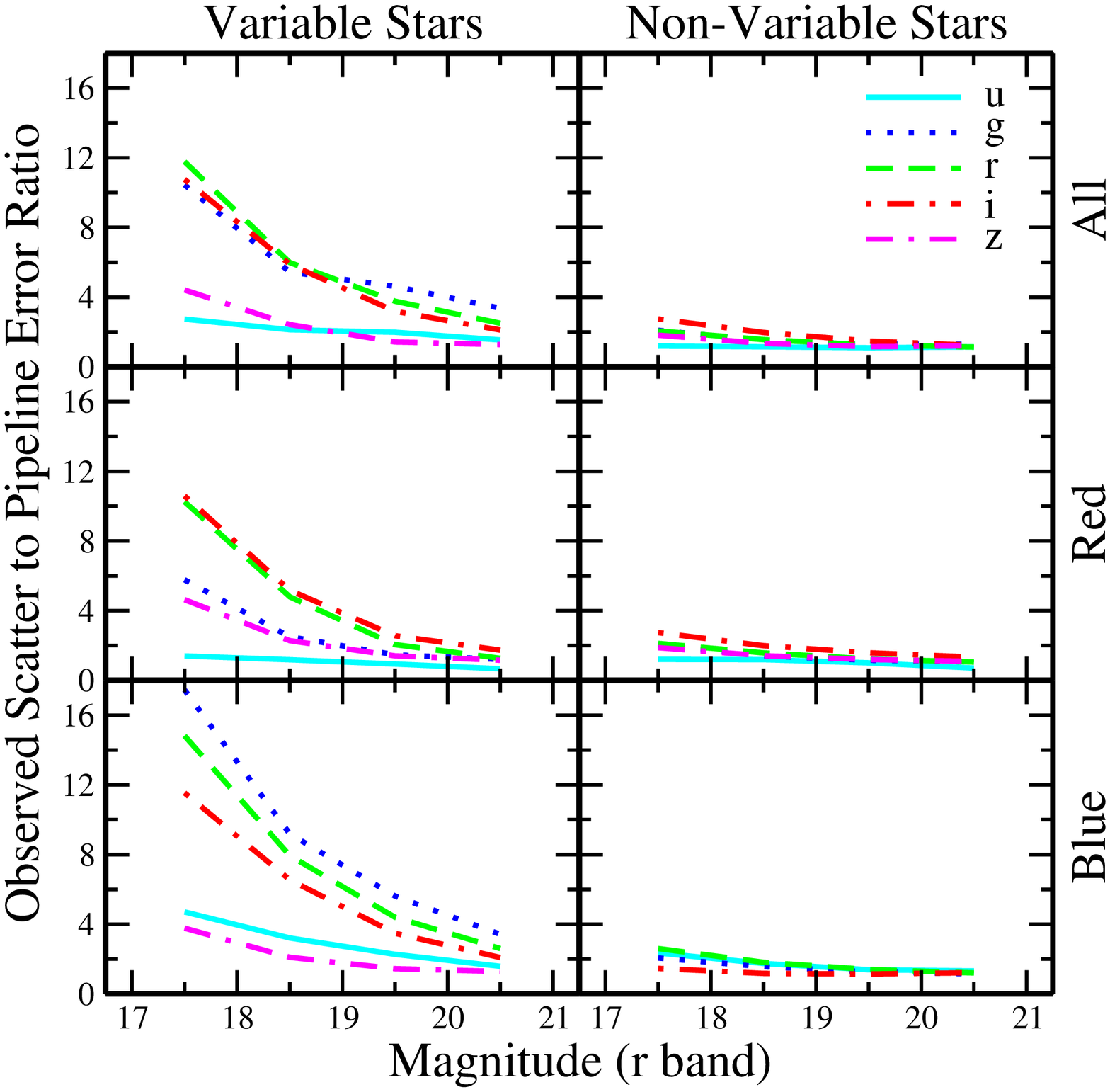}
  \end{center}
  \caption{Same as Figure~\ref{fig:mag_error_model}, but for variable and
    non-variable stars using {\model}.}
  \label{fig:mag_error_model_vary}
\end{figure}

\begin{figure}
  \begin{center}
    \includegraphics[height=6.5in]{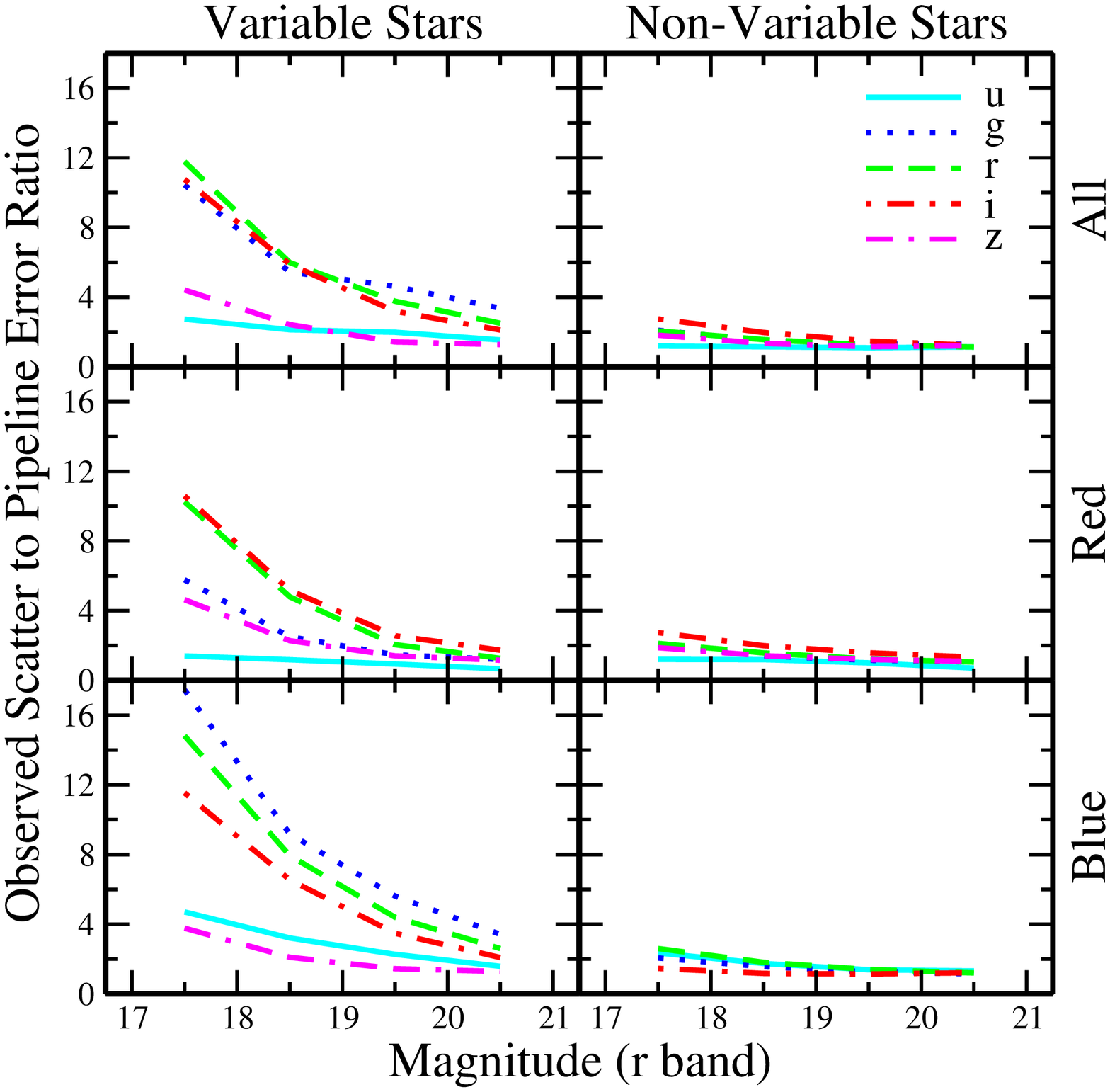}
  \end{center}
  \caption{Same as Figure~\ref{fig:mag_error_model}, but for variable and
    non-variable stars using {\cmodel}.}
  \label{fig:mag_error_cmodel_vary}
\end{figure}

\begin{figure}
  \begin{center}
    \includegraphics[height=6.5in]{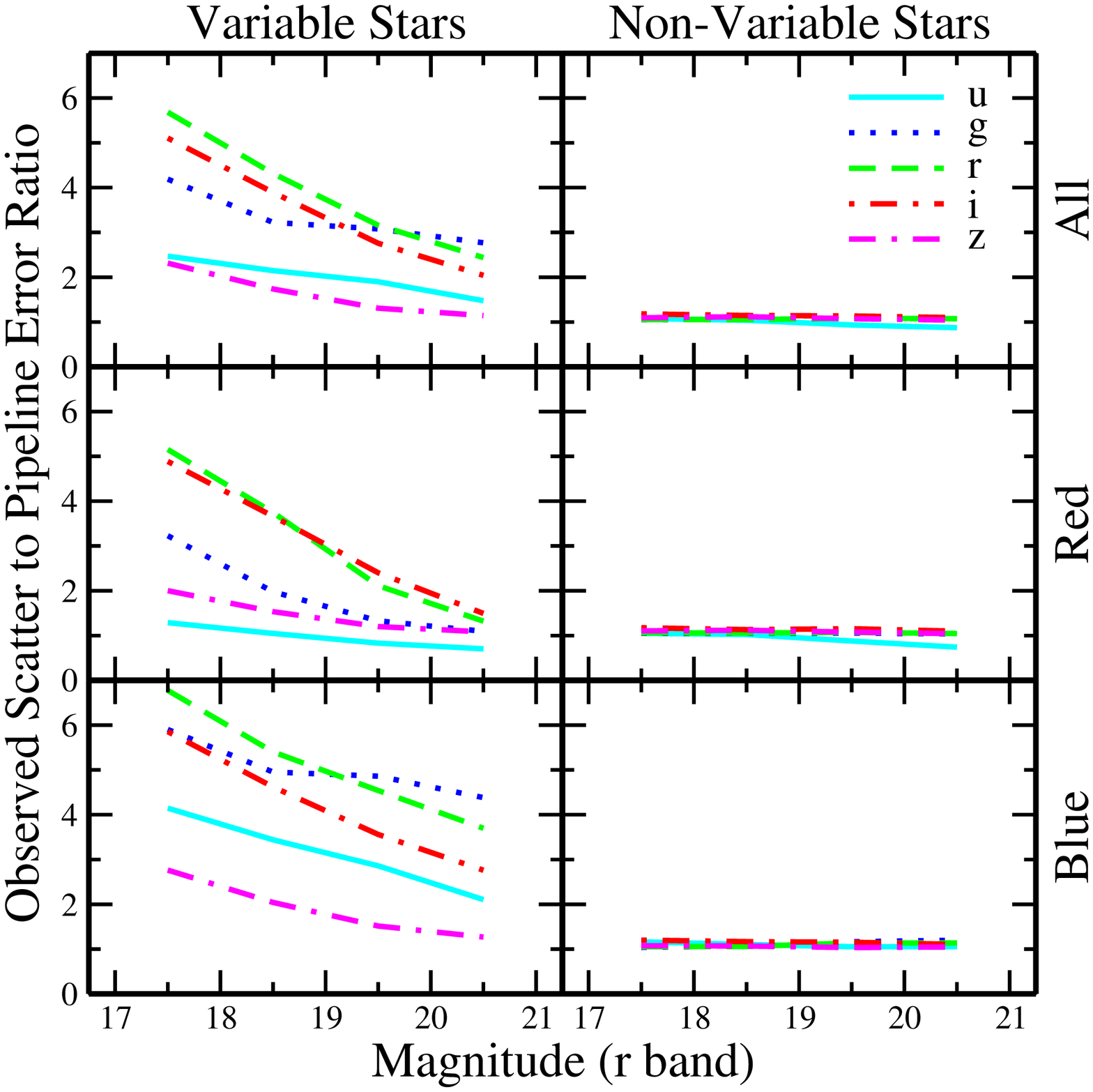}
  \end{center}
  \caption{Same as Figure~\ref{fig:mag_error_model}, but for variable and 
    non-variable stars using {\psf}.}
  \label{fig:mag_error_psf_vary}
\end{figure}

As mentioned in \S\ref{sec:variability}, selecting highly variable objects
by cutting on regression matrix determinant yields 5-10\% of the
total stellar population in any given magnitude/color/aperture bin.  As shown
in Figure~\ref{fig:vary_color_color}, the objects selected by the 
variability cut are primarily located in four regions: the low redshift 
quasar locus, F stars at the blue end of the stellar locus, the K \& M stars 
at the red end of the stellar locus and the blue horizontal branch spur 
extending below the blue end of the stellar locus.  This is consistent with 
where we would expect to find highly variable objects in color-color space,
as well as the results found by \citet{ive00}.  
Figures~\ref{fig:reg_psf_vary} through~\ref{fig:reg_cmodel_vary}
give the variable and non-variable regression matrices as a function of 
magnitude and color for the {\model}, {\cmodel} and {\psf} apertures and
Figures~\ref{fig:mag_error_model_vary} 
through~\ref{fig:mag_error_psf_vary} do the same for the associated 
${\cal R}$ measurements (the values for $\chi^2_{\rm N}$ are in 
Tables~\ref{tab:chi2_vary_star_model} through~\ref{tab:chi2_vary_star_psf}).

As expected by the variability cut, the variable objects show very strong 
covariance between filters.  Contrarily, the non-variable object have nearly 
independent magnitude errors.  The behavior of all three apertures is nearly 
indentical as a function of color and magnitude, as seen in the full stellar 
sample.  This confirms our earlier contention in \S\ref{sec:variability} that 
the covariance between filters for stellar objects seen in 
Figures~\ref{fig:reg_model} through~\ref{fig:reg_petro_psf} was driven 
almost entirely by a small population of highly variable objects in the 
sample, as opposed to the model induced covariance seen in {\model} for 
galaxies.  Further, we can see from the variation of the regression matrices
with color that objects at the blue end of the stellar locus are more highly
variable over the entire optical spectrum than those at the red end of the 
stellar locus.   

Although we did not explicitly select objects with large values of 
$\chi^2_{\rm N}$ with our variability cut, clearly the variable and non-variable
objects have a wide disagreement in $\chi^2_{\rm N}$ for all magnitude and 
colors.  For the latter sub-population, the pipeline errors estimate the 
observed scatter much better than for the entire population (as seen in 
Tables~\ref{tab:chi2_model}, \ref{tab:chi2_cmodel}, and~\ref{tab:chi2_psf}), 
while variable objects display a much larger scatter than the pipeline errors 
would suggest.  Further, while the distribution of $\chi^2_F$ values for the 
non-variable objects is well matched to the expected $\chi^2$ distribution,
the values of $\chi^2_{\rm N}$ for the variable objects are dominated by 
a sizeable fraction of extreme outliers ($\chi^2_F/N > 20$).  This is an 
excellent confirmation that our variability criteria, while perhaps not 
capturing all of the variable objects, does select a sub-population with 
much larger scatter than photon noise would predict.  Futher, as was seen 
with the regression matrices, most of the disagreement with the pipeline 
errors appears to be driven by a relatively small population of highly 
variable objects.  Likewise, the values of $\chi^2_{\rm N}$ for blue objects 
are typically much larger than those for red objects, behavior that is 
consistent regardless of aperture.  This contrasts strongly with the 
non-variable objects, where the color variation is almost nil (as one would 
expect for relatively weak variation of the PSF with color over this range).  
Finally, we can see that blue stellar objects show a stronger disagreement 
relative to red objects at fainter magnitudes than at brighter magnitudes.  
However, this may be an artifact of our selection criteria, which would tend 
to lose red variable objects at brighter magnitudes due to their much lower 
flux in $\uprime$.

\subsection{Isolated vs. Deblended Objects}\label{sec:deblend}

\begin{figure}
  \begin{center}
    \includegraphics[height=6.5in]{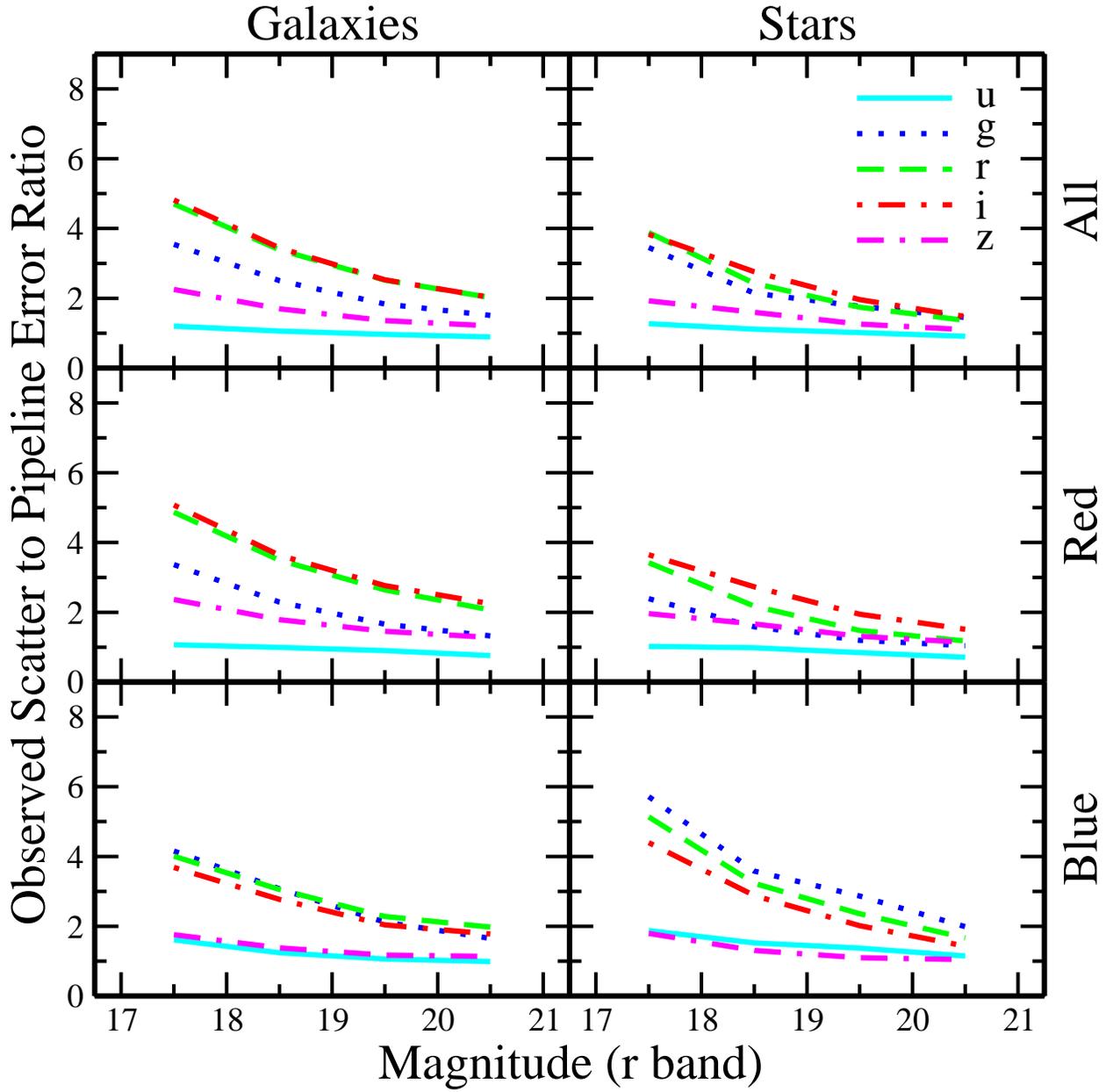}
  \end{center}
  \caption{Same as Figure~\ref{fig:mag_error_model}, but for isolated objects.
  The scaling on the y-axis is the same as in Figure~\ref{fig:mag_error_model}
  to demonstrate the difference in the scatter for blended versus isolated
  objects.}
  \label{fig:mag_error_model_nochild}
\end{figure}

\begin{figure}
  \begin{center}
    \includegraphics[height=6.5in]{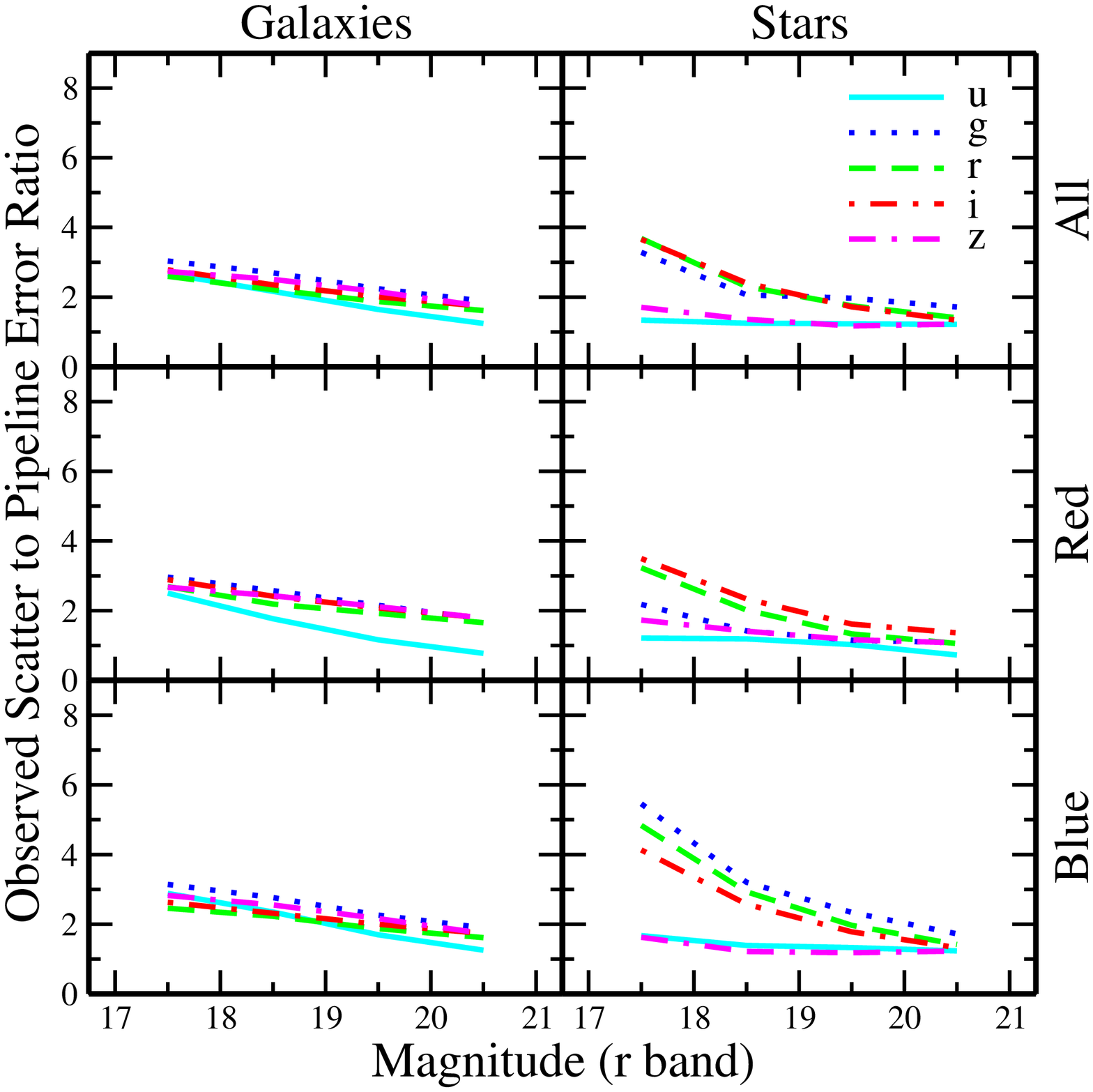}
  \end{center}
  \caption{Same as Figure~\ref{fig:mag_error_model_nochild}, but using 
    {\cmodel}.}
  \label{fig:mag_error_cmodel_nochild}
\end{figure}

\begin{figure}
  \begin{center}
    \includegraphics[height=6.5in]{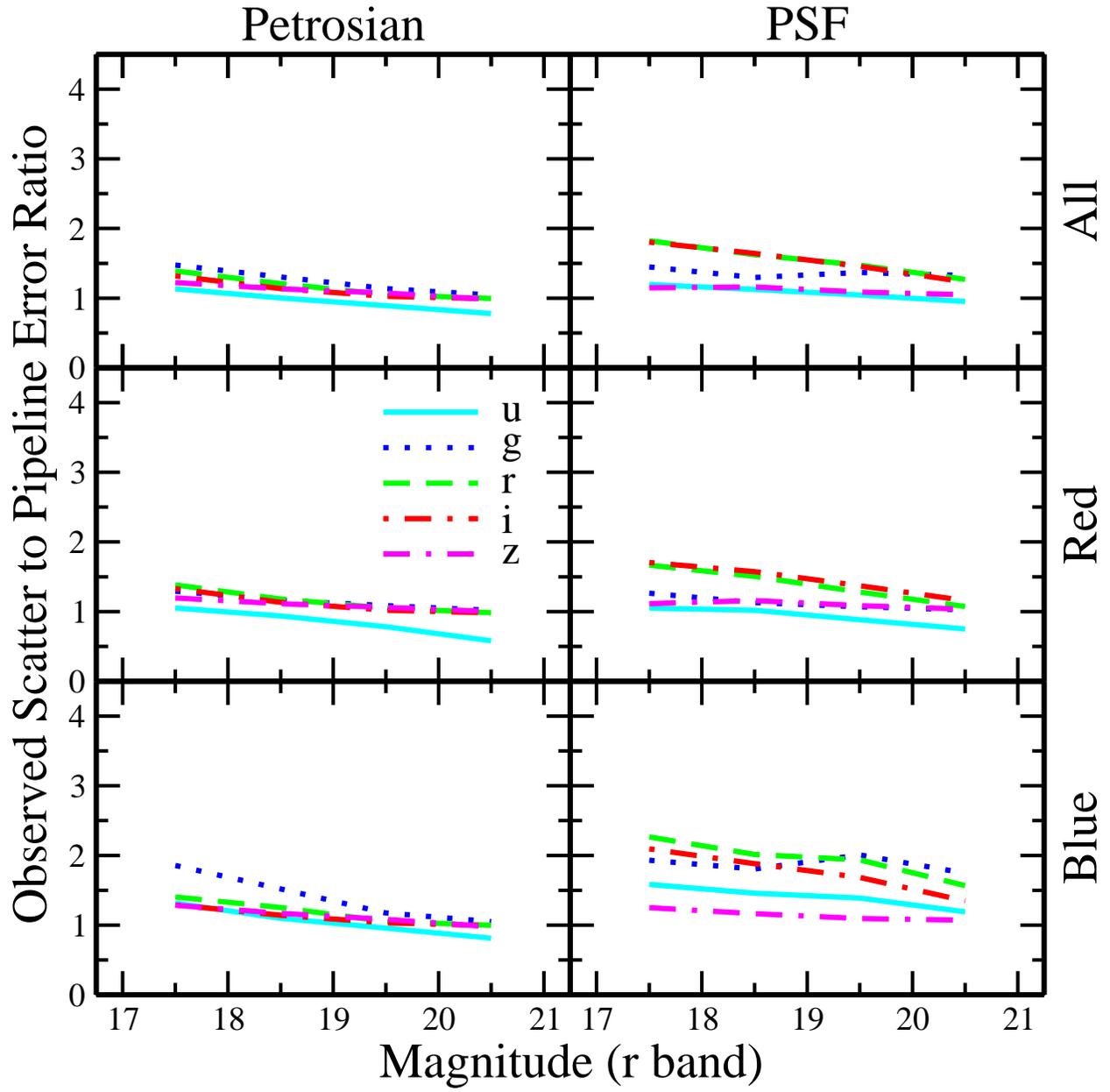}
  \end{center}
  \caption{Same as Figure~\ref{fig:mag_error_model_nochild}, but for galaxies
    using {\petro} and stars using {\psf}.}
  \label{fig:mag_error_petro_psf_nochild}
\end{figure}

Figures~\ref{fig:mag_error_model_nochild} through
\ref{fig:mag_error_petro_psf_nochild} show ${\cal R}$ for the isolated objects.
In all three cases, we keep the scaling on the y-axis identical to those where
we have included the deblended objects (Figures~\ref{fig:mag_error_model}
through~\ref{fig:mag_error_petro_psf}).  As one might expect, the effect of
deblending on magnitude scatter shows up strongest for relatively bright 
galaxies, while faint galaxies and stars at all magnitudes are relatively 
unaffected by whether the object is isolated or part of a blend.  One can see
from Table~\ref{tab:obj_counts} that the fraction of deblended objects in 
a given magnitude/object type bin follows a similar pattern: a large fraction 
of bright galaxies have been deblended, but this ratio shrinks considerably
as galaxies grow fainter; the stellar ratio stays roughly constant with 
magnitude.  The contrast between isolated and blended objects is strongest 
for {\petro}, where {\cal R} is nearly unity for isolated objects at all 
magnitudes, while {\cal R} for the full data set is above 2 at the bright 
end.  {\cmodel} shows similar but less dramatic behavior, while {\model} 
is only changed by 20-30\% at the bright end for isolated versus blended 
objects.

Unlike {\cal R}, the regression matrices for the isolated objects were nearly
identical to the full data set, with individual values of $R_{i,j}$ varying
by less than 5\%.

\section{Error Translation \& Covariance}\label{sec:translate}

With the results of the previous sections, we can develop a prescription
for translating {\photo} errors into observed scatter as well as generating
a covariance matrix.  Since we are primarily interested in applying this 
method to correct color errors, we will concentrate on galaxies using {\model}
and {\petro} and stars using {\psf}.  

The basic relation between the observed scatter $\delta m$ and the pipeline
error $\Delta m$ is 
\begin{equation}
  \delta m \equiv {\cal R} \Delta m,
\end{equation}
where the curves for {\cal R} as a function of aperture, object type, 
magnitude and color are given in Figures~\ref{fig:mag_error_model} through 
\ref{fig:mag_error_petro_psf}.  To model {\cal R}, we will typically need 
two pieces: a term that varies with magnitude to characterize the transition
between errors dominated by modeling fits to those dominated by photon noise
and a constant term representing the floor on the scatter (or equivalently
the gain on the CCDs).  We can approximate these requirements with a simple 
power law plus a constant:
\begin{equation}
  {\cal R}(m) = \left ( \frac{m}{m_0} \right)^\alpha + \beta,
  \label{eq:pipeline2scatter}
\end{equation} 
where $\alpha$, $m_0$ and $\beta$ are a function of filter, aperture and object
type.  For the most accurate results, these parameters should also be a 
function of color, but using the fits ignoring color is sufficient for most
cases.  Table~\ref{tab:pipeline2scatter} presents the fits for the three cases 
mentioned above as well as galaxies using {\cmodel}.  
Equation~\ref{eq:pipeline2scatter} does an excellent job of modeling 
${\cal R}$ for {\model} and {\cmodel} galaxies (with the exception of the
$\uprime$ band in the latter case).  It is a reasonably good approximation
of the variation for {\petro} galaxies, with the same $\uprime$ caveat.  For
{\psf}, the variation of ${\cal R}$ with magnitude is small enough that
we can approximate it best using just the $\beta$ parameter, setting $\alpha$
to zero.  To the extent that we are able to cleanly make the measurements, 
extrapolating Equation~\ref{eq:pipeline2scatter} brightward and faintward of
$17 \le \rprime \le 21$ matches the observed values of ${\cal R}$.

To calculate the covariance matrix, we focus on reproducing the regression
matrix as a function of magnitude and color for a given aperture/object type
combination.  Once the regression matrix is calculated, we can conver it to 
a covariance matrix according to Equation~\ref{eq:mag_covar}.

To handle the variation in the regression matrix as a function of color, we 
use two matrices: the red object regression matrix ($R_{R}$) and the color 
differential matrix ($R_{\rm D} \equiv R_{\rm B} - R_{\rm R}$, where $R_{\rm B}$ 
is the blue object regression matrix).  To produce the regression matrix, we 
combine $R_{\rm R}$ and $R_{\rm D}$ using a sigmoid function:
\begin{equation}
  R = R_{\rm R} + \left ( 1 - 
  \left [ 1 + {\rm exp} \left(-\frac{{\cal C} - {\cal C}_R}
    {\sigma_R} \right) \right]^{-1} \right ) R_{\rm D},
  \label{eq:regression_color}
\end{equation}
where ${\cal C}$ is the color used to separate red and blue objects, 
${\cal C}_R$ is the dividing line between red and blue and $\sigma_R$ controls
the width of the transition.  For our implementation, 
${\cal C} \equiv \uprime - \rprime$ and ${\cal C}_R = 1.8$ as set in 
Equation~\ref{eq:color_sep}.  For {\model} galaxies we set 
$\sigma_R \equiv 0.15$.  The values for $R_{\rm R}$ and $R_{\rm D}$ are given in 
Table~\ref{tab:regression_matrices}.

In addition to setting the amplitude of the off-diagonal elements of $R$ 
according to object color, we also need to take into account the observed
variation of the off-diagonal elements as a function of magnitude as 
described in \S\ref{sec:regression}.  Because objects have intrinsic colors
and the five filters have different depths, we need to model this variation
separately for each element of the regression matrix.  By calculating the 
regression matrix for each unique object and dividing each element by the 
corresponding element from the appropriate regression matrix produced by 
Equation~\ref{eq:regression_color}.  This ratio can be modeled as a function
of $\rprime$ using a simple power law:
\begin{equation}
  {\cal A}_{ij}(\rprime) = \left (\frac{\rprime}{r_{0,ij}}\right)^{A_{ij}}
  \label{eq:regression_amp}
\end{equation}
The values for $r_{0,ij}$ and $A_{ij}$ are given in 
Table~\ref{tab:regression_amp}.  The addition of this term modifies 
Equation~\ref{eq:regression_color} to
\begin{equation}
  R_{ij} = {\cal A}_{ij}(\rprime) \left(R_{{\rm R},ij} + \left ( 1 - 
  \left [ 1 + {\rm exp} \left(-\frac{{\cal C} - {\cal C}_R}
    {\sigma_R} \right) \right]^{-1} \right ) R_{{\rm D},ij} \right),
  \label{eq:regression_full}
\end{equation}
where the diagonal elements of $R$ are set to unity by definition.  

As one might expect from the generally noisy behavior of the $\uprime$ filter,
the variation for the regression elements involving that filter tends to be
much stronger than other filter combinations.  Because of this behavior, the
regression matrices produced by Equation~\ref{eq:regression_full} are 
singular for blue galaxies with $\rprime < 16$.  For these galaxies, using
$\rprime = 16$ should produce a suffciently accurate regression matrix for
most purposes.

\section{Conclusions \& Discussion}\label{sec:conclusions}

In this paper, we have presented an analysis of the observed photometric
covariance for the five SDSS filters drawn from multiple repeat scans of
the southern equatorial stripe.  Given the large number of objects in
the stripe we were able to sub-divide the total sample by magnitude,
object type and color.  Likewise, we looked at the effect of the
standard SDSS apertures ({\model}, {\cmodel}, {\psf}, {\petro}) on the
photometric covariance.  In general we find that the photometric
pipeline ({\photo}) errors under-estimate the observed scatter in the
five filters, although the degree of disagreement was a strong function
of aperture and magnitude.  {\psf} errors were typically under-estimated
by 20-50\%, while the ratio of {\model} and {\cmodel} observed scatter
to {\photo} errors was as large as 6 at bright magnitudes, tailing off
to a ratio of 2 at the faint limit.

The degree of covariance between filters was primarily a function of
object type.  Stellar objects produced similar regression matrices
regardless of aperture.  The amplitude of the off-diagonal elements in
the regression was a weak function of magnitude, typically dropping by
10-20\% from the brightest to the weakest samples.  This variation was
weakest in {\psf}, the preferred aperture for SDSS quasar target
selection.  When variable stellar objects were removed from the sample, 
the correlation between filters was negligible and the pipeline errors
for {\psf} were an excellent match for the observed scatter.

For galaxies, the regression matrix was a strong function of aperture.  
Galaxies observed with {\model} (the preferred aperture for
galaxy colors) were strongly covariant, showing correlations between the
$\gprime$, $\rprime$ and $\iprime$ bands in excess of 70\% for all
colors and magnitudes.  This strong covariance should lead to a large
over-estimation of the color errors if neglected.  However, since the
magnitude errors in {\model} were so strongly under-estimated, the
{\photo} color errors were typically under-estimated by 10-20\% relative
to the observed color scatter, although this did rise to 100\% at the
bright limit.  The scatter at the bright end was also a function of whether
or not the galaxy had been deblended from a larger group.  Bright isolated
galaxies had smaller scatter relative to their deblended counterparts, although
the ratio between the observed scatter and pipeline errors remained large.

Given all of this, we can make the following prescription:\\
$\bullet$ {\bf For Stellar Objects}: Using {\psf} is recommended
for both magnitudes and colors.  In both cases, the pipeline errors match
the observed scatter very well, provided that the object is not intrinsically
variable.  For quasars and variable stars (5-10\% of the total population of
stellar objects), the scatter can be considerably larger than the pipeline 
errors and will tend to be strongly correlated between filters.\\
$\bullet$ {\bf For Galaxy Magnitudes}:  The preferred aperture is 
{\cmodel}, although {\petro} is acceptable for objects bright 
enough to be in the main SDSS galaxy sample ($r < 17.5$).  In either case, 
the pipeline errors can be corrected using Equation~\ref{eq:pipeline2scatter}
with the appropriate parameters from Table~\ref{tab:pipeline2scatter}.\\
$\bullet$ {\bf For Galaxy Colors}: The preferred aperture is {\model}.
At the faint end, the color errors from the pipeline are only slightly 
smaller than the observed scatter, but one should be aware that the ratio
between the observed scatter and the pipeline errors can be larger than 2
for bright objects.  For applications like photometric redshifts, where the 
entire galaxy SED can be important, one should calculatd the full 
magnitude covariance matrix using Equations~\ref{eq:pipeline2scatter} and 
\ref{eq:regression_full}.

Despite the fact that the differences between the observed color errors
in the SDSS and the color errors from the processing pipeline were generally
small, the existence of covariance between magnitude bands is an intrinsic
feature of multi-waveband photometry.  This makes it an issue that
will need to be addressed by nearly all future surveys.  The degree
of precision required for color-based target selection and photometric
redshifts to meet these surveys' science goals will place enormous demands 
on the full photometric system, hardware and software.  Given the strength of 
even the far off-diagonal elements of the {\model} regression matrices, it is 
clear that incorporating the full covariance matrix into the data model for
these surveys will be absolutely necessary to extract the full statistical 
power from these data sets.  

This analysis suggests that future surveys will need to
incorporate large numbers of repeated scans into the early stages of
their survey operations.  This data set will provide an invaluable tool
for empirically checking the ability of the photometric system to
recover the observed covariance for objects as well as excellent means
for testing the efficiency and completeness of target selection,
photometric redshifts and other color-based selection algorithms.
Likewise, it would also serve as an excellent test bed for future
refinements of the photometric pipeline software.

\acknowledgments

The authors would like to thank James Annis, Sebastian Jester, Huan Lin and
Michael Strauss for many useful comments.

Funding for the creation and distribution of the SDSS Archive has been provided
by the Alfred P. Sloan Foundation, the Participating Institutions, the National
Aeronautics and Space Administration, the National Science Foundation, the U.S.
Department of Energy, the Japanese Monbukagakusho, and the Max Planck Society. 
The SDSS Web site is http://www.sdss.org/.

The SDSS is managed by the Astrophysical Research Consortium (ARC) for the 
Participating Institutions.  The Participating Institutions are The University 
of Chicago, Fermilab, the Institute for Advanced Study, the Japan Participation
Group, The Johns Hopkins University, the Korean Scientist Group, Los Alamos 
National Laboratory, the Max-Planck-Institute for Astronomy (MPIA), the 
Max-Planck-Institute for Astrophysics (MPA), New Mexico State University, 
University of Pittsburgh,  University of Portsmouth,  Princeton University, 
the United States Naval Observatory, and the University of Washington.

\clearpage

\clearpage

\clearpage

\clearpage

\clearpage

\clearpage

\clearpage

\clearpage

\clearpage

\clearpage

\clearpage

\clearpage

\clearpage

\clearpage

\clearpage

\clearpage

\clearpage

\clearpage

\clearpage

\clearpage

\begin{deluxetable}{ccccccccc}
\tabletypesize{\footnotesize} 
\tablecolumns{9}
\tablewidth{0pt}
\tablecaption{Unique objects per magnitude and color bin for objects with 
at least 10 epochs.}
\tablehead{
\colhead{} & \colhead{} & \multicolumn{3}{c}{Galaxies} & \colhead{} & 
\multicolumn{3}{c}{Stars}\\
\cline{3-5} \cline{7-9} \\
\colhead{Aperture} & \colhead{Mag. Limit} & 
\colhead{All} & \colhead{Red} & \colhead{Blue} & \colhead{} &
\colhead{All} & \colhead{Red} & \colhead{Blue}
}
\startdata
All {\model} & $17 < \rprime < 18$ & 14505 & 11744 & 2761 & & 52293 & 40258 & 12035 \\
       & $18 < \rprime < 19$ & 48411 & 37301 & 11110 & & 68046 & 53566 & 14480 \\
       & $19 < \rprime < 20$ & 141876 & 98927 & 42949 & & 99440 & 75698 & 23742 \\
       & $20 < \rprime < 21$ & 335314 & 196027 & 140287 & & 140187 & 98770 & 41417 \\\hline
Isolated {\model} & $17 < \rprime < 18$ & 7535 & 5972 & 1563 & & 37266 & 28791 & 8475\\
                  & $18 < \rprime < 19$ & 32204 & 24385 & 7819 & & 53144 & 42539 & 10605\\
                  & $19 < \rprime < 20$ & 106569 & 72003 & 34566 & & 78840 & 60630 & 18210\\
                  & $20 < \rprime < 21$ & 283497 & 157933 & 125564 & & 117163 & 81642 & 35521\\\hline\hline
All {\cmodel} & $17 < \rprime < 18$ & 14631 & 8943 & 5688 & & 52294 & 39629 & 12665\\
       & $18 < \rprime < 19$ & 48993 & 17294 & 31699 & & 68046 & 48462 & 19584\\
       & $19 < \rprime < 20$ & 143509 & 17339 & 126170 & & 99442 & 39389 & 60053\\
       & $20 < \rprime < 21$ & 338506 & 10423 & 328083 & & 140176 & 16687 & 123489\\\hline
Isolated {\cmodel} & $17 < \rprime < 18$ & 7629 & 4364 & 3265 & & 37268 & 28307 & 8961\\
       & $18 < \rprime < 19$ & 32714 & 10532 & 22182 & & 53143 & 38364 & 14779\\
       & $19 < \rprime < 20$ & 108011 & 10962 & 97049 & & 78843 & 30968 & 47875\\
       & $20 < \rprime < 21$ & 285944 & 6991 & 278953 & & 117157 & 13032 & 104125\\\hline\hline
All {\petro} & $17 < \rprime < 18$ & 13588 & 10210 & 3378 &  & \nodata & \nodata & \nodata\\
       & $18 < \rprime < 19$ & 46108 & 31527 & 14581 & &  \nodata & \nodata & \nodata\\
       & $19 < \rprime < 20$ & 136891 & 77778 & 59113 & &  \nodata & \nodata & \nodata\\
       & $20 < \rprime < 21$ & 312519 & 126860 & 185659 & &  \nodata & \nodata & \nodata\\\hline
Isolated {\petro} & $17 < \rprime < 18$ & 6884 & 4941 & 1943 &  & \nodata & \nodata & \nodata\\
       & $18 < \rprime < 19$ & 30400 & 20114 & 10286 & &  \nodata & \nodata & \nodata\\
       & $19 < \rprime < 20$ & 101991 & 55025 & 46966 & &  \nodata & \nodata & \nodata\\
       & $20 < \rprime < 21$ & 260567 & 99102 & 161465 & &  \nodata & \nodata & \nodata\\\hline\hline
All {\psf} & $17 < \rprime < 18$ & \nodata & \nodata & \nodata &  & 52233 & 40070 & 12163 \\ 
       & $18 < \rprime < 19$ & \nodata & \nodata & \nodata & &  67857 & 53280 &
14577 \\
       & $19 < \rprime < 20$ & \nodata & \nodata & \nodata & &  98537 & 74636 & 23901 \\
       & $20 < \rprime < 21$ & \nodata & \nodata & \nodata & &  137671 & 96214 & 41457 \\\hline
Isolated {\psf} & $17 < \rprime < 18$ & \nodata & \nodata & \nodata &  & 37207 & 28645 & 8562 \\ 
       & $18 < \rprime < 19$ & \nodata & \nodata & \nodata & &  53001 & 42312 &
10689 \\
       & $19 < \rprime < 20$ & \nodata & \nodata & \nodata & &  78123 & 59804 & 18319 \\
       & $20 < \rprime < 21$ & \nodata & \nodata & \nodata & &  114798 & 79362 & 35436 \\
\enddata
\label{tab:obj_counts}
\end{deluxetable}

\clearpage

\begin{deluxetable}{cccccccc}
\tablecolumns{8}
\tablewidth{0pt}
\tablecaption{$\chi^2_{\rm N}$ values for each band as a function of object 
type, magnitude and color for {\model}.}
\tablehead{
\colhead{Object Type} & \colhead{Magnitude Limit} & \colhead{Color Cut} &
\colhead{$\uprime$} & \colhead{$\gprime$} & \colhead{$\rprime$} &
\colhead{$\iprime$} & \colhead{$\zprime$} 
}
\startdata
 Galaxy  & $17 < \rprime < 18$ & All & 1.80 & 26.4 & 46.1 & 48.5 & 10.7 \\
         &                     & Red & 1.37 & 22.3 & 45.8 & 50.8 & 11.6 \\
         &                     & Blue & 3.63 & 43.4 & 47.8 & 38.8 & 6.69 \\
\hline
         & $18 < \rprime < 19$ & All & 1.22 & 9.07 & 17.8 & 19.3 & 4.33 \\
         &                     & Red & 1.04 & 7.66 & 18.8 & 21.1 & 4.76 \\
         &                     & Blue & 1.81 & 15.1 & 16.1 & 13.6 & 2.67 \\
\hline
         & $19 < \rprime < 20$ & All & 0.95 & 4.01 & 7.97 & 8.33 & 2.16 \\
         &                     & Red & 0.80 & 3.12 & 8.29 & 9.39 & 2.39 \\
         &                     & Blue & 1.22 & 5.83 & 6.78 & 5.41 & 1.52 \\
\hline
         & $20 < \rprime < 21$ & All & 0.77 & 2.48 & 4.51 & 4.61 & 1.55 \\
         &                     & Red & 0.54 & 1.90 & 4.70 & 5.76 & 1.76 \\
         &                     & Blue & 0.97 & 3.02 & 4.27 & 3.46 & 1.32 \\
\hline \hline
 Star    & $17 < \rprime < 18$ & All & 1.72 & 13.3 & 14.4 & 15.9 & 4.82 \\
         &                     & Red & 1.09 & 6.74 & 10.9 & 13.9 & 5.62 \\
         &                     & Blue & 3.35 & 29.2 & 21.7 & 17.4 & 3.49 \\
\hline
         & $18 < \rprime < 19$ & All & 1.33 & 5.55 & 6.20 & 7.78 & 2.77 \\
         &                     & Red & 0.97 & 2.92 & 4.56 & 7.41 & 2.86 \\
         &                     & Blue & 2.49 & 14.1 & 10.4 & 8.69 & 1.87 \\
\hline
         & $19 < \rprime < 20$ & All & 1.06 & 3.60 & 3.18 & 4.13 & 1.68 \\
         &                     & Red & 0.70 & 1.54 & 2.17 & 4.01 & 1.83 \\
         &                     & Blue & 1.95 & 8.97 & 5.92 & 4.18 & 1.24 \\
\hline
         & $20 < \rprime < 21$ & All & 0.82 & 2.21 & 1.96 & 2.32 & 1.24 \\
         &                     & Red & 0.49 & 1.13 & 1.47 & 2.36 & 1.32 \\
         &                     & Blue & 1.37 & 4.57 & 3.10 & 2.22 & 1.10 \\
\enddata
\label{tab:chi2_model}
\end{deluxetable}

\clearpage

\begin{deluxetable}{cccccccc}
\tablecolumns{8}
\tablewidth{0pt}
\tablecaption{Same as Table~\ref{tab:chi2_model}, but for {\cmodel}.}
\tablehead{
\colhead{Object Type} & \colhead{Magnitude Limit} & \colhead{Color Cut} &
\colhead{$\uprime$} & \colhead{$\gprime$} & \colhead{$\rprime$} &
\colhead{$\iprime$} & \colhead{$\zprime$} 
}
\startdata
 Galaxy  & $17 < \rprime < 18$ & All & 7.36 & 21.4 & 27.6 & 31.6 & 13.1 \\
         &                     & Red & 6.53 & 22.7 & 33.8 & 40.2 & 14.9 \\
         &                     & Blue & 8.67 & 19.3 & 17.6 & 17.5 & 10.3 \\
\hline
         & $18 < \rprime < 19$ & All & 4.67 & 10.4 & 9.33 & 11.4 & 7.31 \\
         &                     & Red & 3.06 & 9.99 & 11.8 & 15.6 & 7.79 \\
         &                     & Blue & 5.47 & 10.5 & 8.50 & 9.26 & 6.97 \\
\hline
         & $19 < \rprime < 20$ & All & 2.60 & 5.89 & 4.86 & 5.81 & 4.81 \\
         &                     & Red & 1.23 & 5.57 & 5.89 & 7.81 & 4.82 \\
         &                     & Blue & 2.77 & 5.94 & 4.63 & 5.34 & 4.81 \\
\hline
         & $20 < \rprime < 21$ & All & 1.49 & 3.90 & 2.83 & 3.34 & 2.99 \\
         &                     & Red & 0.53 & 3.24 & 3.29 & 4.72 & 3.12 \\
         &                     & Blue & 1.52 & 3.98 & 2.97 & 3.39 & 3.00 \\
\hline \hline
 Star    & $17 < \rprime < 18$ & All & 1.94 & 12.1 & 12.8 & 15.5 & 4.32 \\
         &                     & Red & 1.49 & 6.28 & 11.4 & 14.3 & 4.83 \\
         &                     & Blue & 2.71 & 26.6 & 19.6 & 15.9 & 3.02 \\
\hline
         & $18 < \rprime < 19$ & All & 1.65 & 5.50 & 6.36 & 7.09 & 2.12 \\
         &                     & Red & 1.38 & 2.36 & 4.24 & 6.00 & 2.34 \\
         &                     & Blue & 2.04 & 11.2 & 8.53 & 7.28 & 1.69 \\
\hline
         & $19 < \rprime < 20$ & All & 1.55 & 4.42 & 3.27 & 3.13 & 1.44 \\
         &                     & Red & 1.00 & 1.51 & 1.92 & 3.14 & 1.45 \\
         &                     & Blue & 1.73 & 5.83 & 3.82 & 3.29 & 1.39 \\
\hline
         & $20 < \rprime < 21$ & All & 1.49 & 3.18 & 2.12 & 1.97 & 1.49 \\
         &                     & Red & 0.49 & 1.23 & 1.21 & 2.02 & 1.21 \\
         &                     & Blue & 1.49 & 3.02 & 2.02 & 1.82 & 1.50 \\
\enddata
\label{tab:chi2_cmodel}
\end{deluxetable}

\clearpage

\begin{deluxetable}{cccccccc}
\tablecolumns{8}
\tablewidth{0pt}
\tablecaption{Same as Table~\ref{tab:chi2_model}, but for {\petro}.}
\tablehead{
\colhead{Object Type} & \colhead{Magnitude Limit} & \colhead{Color Cut} &
\colhead{$\uprime$} & \colhead{$\gprime$} & \colhead{$\rprime$} &
\colhead{$\iprime$} & \colhead{$\zprime$} 
}
\startdata
 Galaxy  & $17 < \rprime < 18$ & All & 1.33 & 4.55 & 6.03 & 6.36 & 2.61 \\
         &                     & Red & 1.16 & 3.88 & 6.13 & 7.13 & 2.83 \\
         &                     & Blue & 1.84 & 6.55 & 5.76 & 4.06 & 1.96 \\
\hline
         & $18 < \rprime < 19$ & All & 1.02 & 2.21 & 2.15 & 2.10 & 1.46 \\
         &                     & Red & 0.88 & 1.73 & 2.15 & 2.53 & 1.48 \\
         &                     & Blue & 1.23 & 2.80 & 2.21 & 1.83 & 1.42 \\
\hline
         & $19 < \rprime < 20$ & All & 0.79 & 1.39 & 1.26 & 1.21 & 1.15 \\
         &                     & Red & 0.58 & 1.22 & 1.27 & 1.21 & 1.12 \\
         &                     & Blue & 0.92 & 1.50 & 1.27 & 1.18 & 1.16 \\
\hline
         & $20 < \rprime < 21$ & All & 0.59 & 1.11 & 1.04 & 1.03 & 0.95 \\
         &                     & Red & 0.31 & 1.04 & 1.05 & 1.03 & 1.00 \\
         &                     & Blue & 0.65 & 1.14 & 1.05 & 1.02 & 0.93 \\
\enddata
\label{tab:chi2_petro}
\end{deluxetable}

\begin{deluxetable}{cccccccc}
\tablecolumns{8}
\tablewidth{0pt}
\tablecaption{Same as Table~\ref{tab:chi2_model}, but for {\psf}.}
\tablehead{
\colhead{Object Type} & \colhead{Magnitude Limit} & \colhead{Color Cut} &
\colhead{$\uprime$} & \colhead{$\gprime$} & \colhead{$\rprime$} &
\colhead{$\iprime$} & \colhead{$\zprime$} 
}
\startdata
 Star    & $17 < \rprime < 18$ & All & 1.40 & 1.84 & 2.62 & 2.64 & 1.38 \\
         &                     & Red & 1.13 & 1.51 & 2.33 & 2.42 & 1.39 \\
         &                     & Blue & 2.35 & 3.19 & 3.93 & 3.54 & 1.57 \\
\hline
         & $18 < \rprime < 19$ & All & 1.30 & 1.64 & 2.24 & 2.36 & 1.34 \\
         &                     & Red & 1.05 & 1.29 & 1.88 & 1.93 & 1.33 \\
         &                     & Blue & 2.18 & 3.27 & 3.69 & 3.19 & 1.42 \\
\hline
         & $19 < \rprime < 20$ & All & 1.13 & 1.96 & 2.01 & 1.95 & 1.17 \\
         &                     & Red & 0.76 & 1.16 & 1.41 & 1.65 & 1.20 \\
         &                     & Blue & 2.03 & 4.26 & 3.78 & 2.76 & 1.23 \\
\hline
         & $20 < \rprime < 21$ & All & 0.86 & 1.72 & 1.60 & 1.50 & 1.10 \\
         &                     & Red & 0.54 & 1.09 & 1.18 & 1.33 & 1.09 \\
         &                     & Blue & 1.52 & 3.60 & 2.77 & 1.97 & 1.16 \\
\enddata
\label{tab:chi2_psf}
\end{deluxetable}

\clearpage

\begin{deluxetable}{ccccccc}
\tablecolumns{7}
\tablewidth{0pt}
\tablecaption{Color errors for galaxies using {\model} as a function of 
magnitude limit and color.  See \S\ref{sec:color_errors} for descriptions of 
``Proper Error'', ``Naive Error'', ``{\tt photo} Error'' and 
``Observed Error''.}
\tablehead{
\colhead{Color Cut} & \colhead{Magnitude Limit} & \colhead{Color} &
\colhead{Proper Error} & \colhead{Naive Error} & \colhead{{\tt photo} Error} &
\colhead{Observed Error} 
}
     
\startdata
All & $17 < \rprime < 18$ & $\uprime - \gprime$ & 0.149 & 0.179 & 0.124 & 0.171 \\
    &                     & $\gprime - \rprime$ & 0.037 & 0.088 & 0.016 & 0.033 \\
    &                     & $\rprime - \iprime$ & 0.026 & 0.089 & 0.013 & 0.024 \\
    &                     & $\iprime - \zprime$ & 0.038 & 0.095 & 0.023 & 0.037 \\
\hline
    & $18 < \rprime < 19$ & $\uprime - \gprime$ & 0.195 & 0.224 & 0.192 & 0.273 \\
    &                     & $\gprime - \rprime$ & 0.045 & 0.099 & 0.031 & 0.042 \\
    &                     & $\rprime - \iprime$ & 0.033 & 0.100 & 0.023 & 0.030 \\
    &                     & $\iprime - \zprime$ & 0.051 & 0.110 & 0.043 & 0.054 \\
\hline
    & $19 < \rprime < 20$ & $\uprime - \gprime$ & 0.279 & 0.308 & 0.302 & 0.350 \\
    &                     & $\gprime - \rprime$ & 0.070 & 0.124 & 0.060 & 0.071 \\
    &                     & $\rprime - \iprime$ & 0.047 & 0.125 & 0.044 & 0.047 \\
    &                     & $\iprime - \zprime$ & 0.088 & 0.148 & 0.086 & 0.102 \\
\hline
    & $20 < \rprime < 21$ & $\uprime - \gprime$ & 0.391 & 0.425 & 0.459 & 0.427 \\
    &                     & $\gprime - \rprime$ & 0.125 & 0.193 & 0.121 & 0.139 \\
    &                     & $\rprime - \iprime$ & 0.091 & 0.195 & 0.091 & 0.093 \\
    &                     & $\iprime - \zprime$ & 0.173 & 0.247 & 0.177 & 0.209 \\
\hline \hline
Red & $17 < \rprime < 18$ & $\uprime - \gprime$ & 0.151 & 0.175 & 0.139 & 0.185 \\
    &                     & $\gprime - \rprime$ & 0.034 & 0.086 & 0.016 & 0.033 \\
    &                     & $\rprime - \iprime$ & 0.025 & 0.088 & 0.013 & 0.023 \\
    &                     & $\iprime - \zprime$ & 0.036 & 0.093 & 0.022 & 0.034 \\
\hline
    & $18 < \rprime < 19$ & $\uprime - \gprime$ & 0.245 & 0.266 & 0.251 & 0.304 \\
    &                     & $\gprime - \rprime$ & 0.043 & 0.099 & 0.032 & 0.044 \\
    &                     & $\rprime - \iprime$ & 0.031 & 0.100 & 0.022 & 0.029 \\
    &                     & $\iprime - \zprime$ & 0.047 & 0.108 & 0.040 & 0.048 \\
\hline
    & $19 < \rprime < 20$ & $\uprime - \gprime$ & 0.360 & 0.377 & 0.410 & 0.401 \\
    &                     & $\gprime - \rprime$ & 0.070 & 0.121 & 0.064 & 0.076 \\
    &                     & $\rprime - \iprime$ & 0.044 & 0.122 & 0.041 & 0.044 \\
    &                     & $\iprime - \zprime$ & 0.078 & 0.140 & 0.076 & 0.084 \\
\hline
    & $20 < \rprime < 21$ & $\uprime - \gprime$ & 0.481 & 0.495 & 0.645 & 0.488 \\
    &                     & $\gprime - \rprime$ & 0.132 & 0.176 & 0.130 & 0.151 \\
    &                     & $\rprime - \iprime$ & 0.078 & 0.179 & 0.078 & 0.080 \\
    &                     & $\iprime - \zprime$ & 0.140 & 0.218 & 0.143 & 0.159 \\
\hline \hline \tablebreak
Blue & $17 < \rprime < 18$ & $\uprime - \gprime$ & 0.117 & 0.153 & 0.072 & 0.089 \\
    &                     & $\gprime - \rprime$ & 0.036 & 0.098 & 0.015 & 0.035 \\
    &                     & $\rprime - \iprime$ & 0.028 & 0.097 & 0.015 & 0.028 \\
    &                     & $\iprime - \zprime$ & 0.052 & 0.106 & 0.034 & 0.047 \\
\hline
    & $18 < \rprime < 19$ & $\uprime - \gprime$ & 0.128 & 0.167 & 0.112 & 0.136 \\
    &                     & $\gprime - \rprime$ & 0.040 & 0.101 & 0.026 & 0.038 \\
    &                     & $\rprime - \iprime$ & 0.034 & 0.101 & 0.026 & 0.033 \\
    &                     & $\iprime - \zprime$ & 0.069 & 0.116 & 0.059 & 0.071 \\
\hline
    & $19 < \rprime < 20$ & $\uprime - \gprime$ & 0.212 & 0.247 & 0.210 & 0.234 \\
    &                     & $\gprime - \rprime$ & 0.060 & 0.127 & 0.053 & 0.061 \\
    &                     & $\rprime - \iprime$ & 0.052 & 0.128 & 0.052 & 0.053 \\
    &                     & $\iprime - \zprime$ & 0.117 & 0.164 & 0.117 & 0.129 \\
\hline
    & $20 < \rprime < 21$ & $\uprime - \gprime$ & 0.342 & 0.380 & 0.363 & 0.366 \\
    &                     & $\gprime - \rprime$ & 0.116 & 0.208 & 0.114 & 0.124 \\
    &                     & $\rprime - \iprime$ & 0.103 & 0.212 & 0.109 & 0.105 \\
    &                     & $\iprime - \zprime$ & 0.224 & 0.289 & 0.228 & 0.249 \\
\hline \hline
\enddata
\label{tab:color_error_model_gal}
\end{deluxetable}

\clearpage

\begin{deluxetable}{ccccccc}
\tablecolumns{7}
\tablewidth{0pt}
\tablecaption{Same as Table~\ref{tab:color_error_model_gal} but for stars
using {\model}.}
\tablehead{
\colhead{Color Cut} & \colhead{Magnitude Limit} & \colhead{Color} &
\colhead{Proper Error} & \colhead{Naive Error} & \colhead{{\tt photo} Error} &
\colhead{Observed Error} 
}
\startdata
All & $17 < \rprime < 18$ & $\uprime - \gprime$ & 0.074 & 0.092 & 0.065 & 0.133 \\
    &                     & $\gprime - \rprime$ & 0.032 & 0.034 & 0.012 & 0.025 \\
    &                     & $\rprime - \iprime$ & 0.024 & 0.034 & 0.009 & 0.025 \\
    &                     & $\iprime - \zprime$ & 0.031 & 0.036 & 0.013 & 0.032 \\
\hline
    & $18 < \rprime < 19$ & $\uprime - \gprime$ & 0.131 & 0.148 & 0.125 & 0.269 \\
    &                     & $\gprime - \rprime$ & 0.036 & 0.037 & 0.020 & 0.032 \\
    &                     & $\rprime - \iprime$ & 0.027 & 0.037 & 0.014 & 0.028 \\
    &                     & $\iprime - \zprime$ & 0.036 & 0.040 & 0.021 & 0.041 \\
\hline
    & $19 < \rprime < 20$ & $\uprime - \gprime$ & 0.197 & 0.225 & 0.214 & 0.372 \\
    &                     & $\gprime - \rprime$ & 0.052 & 0.047 & 0.036 & 0.051 \\
    &                     & $\rprime - \iprime$ & 0.036 & 0.045 & 0.024 & 0.037 \\
    &                     & $\iprime - \zprime$ & 0.047 & 0.052 & 0.036 & 0.063 \\
\hline
    & $20 < \rprime < 21$ & $\uprime - \gprime$ & 0.298 & 0.327 & 0.354 & 0.423 \\
    &                     & $\gprime - \rprime$ & 0.081 & 0.073 & 0.069 & 0.094 \\
    &                     & $\rprime - \iprime$ & 0.054 & 0.068 & 0.047 & 0.060 \\
    &                     & $\iprime - \zprime$ & 0.074 & 0.082 & 0.069 & 0.132 \\
\hline \hline
Red & $17 < \rprime < 18$ & $\uprime - \gprime$ & 0.093 & 0.096 & 0.089 & 0.150 \\
    &                     & $\gprime - \rprime$ & 0.026 & 0.030 & 0.011 & 0.025 \\
    &                     & $\rprime - \iprime$ & 0.022 & 0.031 & 0.009 & 0.023 \\
    &                     & $\iprime - \zprime$ & 0.029 & 0.035 & 0.013 & 0.030 \\
\hline
    & $18 < \rprime < 19$ & $\uprime - \gprime$ & 0.201 & 0.202 & 0.203 & 0.302 \\
    &                     & $\gprime - \rprime$ & 0.032 & 0.032 & 0.020 & 0.033 \\
    &                     & $\rprime - \iprime$ & 0.026 & 0.033 & 0.014 & 0.027 \\
    &                     & $\iprime - \zprime$ & 0.033 & 0.037 & 0.019 & 0.037 \\
\hline
    & $19 < \rprime < 20$ & $\uprime - \gprime$ & 0.377 & 0.378 & 0.450 & 0.434 \\
    &                     & $\gprime - \rprime$ & 0.050 & 0.040 & 0.042 & 0.054 \\
    &                     & $\rprime - \iprime$ & 0.032 & 0.040 & 0.024 & 0.035 \\
    &                     & $\iprime - \zprime$ & 0.042 & 0.046 & 0.031 & 0.051 \\
\hline
    & $20 < \rprime < 21$ & $\uprime - \gprime$ & 0.535 & 0.535 & 0.758 & 0.506 \\
    &                     & $\gprime - \rprime$ & 0.093 & 0.065 & 0.091 & 0.107 \\
    &                     & $\rprime - \iprime$ & 0.050 & 0.060 & 0.045 & 0.054 \\
    &                     & $\iprime - \zprime$ & 0.061 & 0.069 & 0.056 & 0.088 \\
\hline \hline \tablebreak
Blue & $17 < \rprime < 18$ & $\uprime - \gprime$ & 0.045 & 0.067 & 0.031 & 0.044 \\
    &                     & $\gprime - \rprime$ & 0.030 & 0.042 & 0.010 & 0.024 \\
    &                     & $\rprime - \iprime$ & 0.024 & 0.041 & 0.009 & 0.027 \\
    &                     & $\iprime - \zprime$ & 0.032 & 0.041 & 0.017 & 0.036 \\
\hline
    & $18 < \rprime < 19$ & $\uprime - \gprime$ & 0.069 & 0.093 & 0.054 & 0.068 \\
    &                     & $\gprime - \rprime$ & 0.032 & 0.047 & 0.015 & 0.029 \\
    &                     & $\rprime - \iprime$ & 0.029 & 0.047 & 0.015 & 0.032 \\
    &                     & $\iprime - \zprime$ & 0.045 & 0.055 & 0.034 & 0.049 \\
\hline
    & $19 < \rprime < 20$ & $\uprime - \gprime$ & 0.110 & 0.145 & 0.097 & 0.121 \\
    &                     & $\gprime - \rprime$ & 0.041 & 0.063 & 0.026 & 0.040 \\
    &                     & $\rprime - \iprime$ & 0.038 & 0.064 & 0.029 & 0.041 \\
    &                     & $\iprime - \zprime$ & 0.079 & 0.091 & 0.074 & 0.088 \\
\hline
    & $20 < \rprime < 21$ & $\uprime - \gprime$ & 0.192 & 0.227 & 0.186 & 0.234 \\
    &                     & $\gprime - \rprime$ & 0.061 & 0.091 & 0.052 & 0.064 \\
    &                     & $\rprime - \iprime$ & 0.064 & 0.095 & 0.059 & 0.069 \\
    &                     & $\iprime - \zprime$ & 0.165 & 0.177 & 0.163 & 0.192\\
\hline \hline
\enddata
\label{tab:color_error_model_star}
\end{deluxetable}

\clearpage

\begin{deluxetable}{ccccccc}
\tablecolumns{7}
\tablewidth{0pt}
\tablecaption{Same as Table~\ref{tab:color_error_model_gal} but for galaxies
using {\cmodel}.}
\tablehead{
\colhead{Color Cut} & \colhead{Magnitude Limit} & \colhead{Color} &
\colhead{Proper Error} & \colhead{Naive Error} & \colhead{{\tt photo} Error} &
\colhead{Observed Error} 
}
\startdata
All & $17 < \rprime < 18$ & $\uprime - \gprime$ & 0.385 & 0.389 & 0.143 & 0.329 \\
    &                     & $\gprime - \rprime$ & 0.056 & 0.063 & 0.015 & 0.056 \\
    &                     & $\rprime - \iprime$ & 0.041 & 0.066 & 0.012 & 0.040 \\
    &                     & $\iprime - \zprime$ & 0.076 & 0.093 & 0.024 & 0.080 \\
\hline
    & $18 < \rprime < 19$ & $\uprime - \gprime$ & 0.525 & 0.528 & 0.243 & 0.407 \\
    &                     & $\gprime - \rprime$ & 0.084 & 0.067 & 0.029 & 0.092 \\
    &                     & $\rprime - \iprime$ & 0.065 & 0.078 & 0.024 & 0.058 \\
    &                     & $\iprime - \zprime$ & 0.134 & 0.141 & 0.050 & 0.136 \\
\hline
    & $19 < \rprime < 20$ & $\uprime - \gprime$ & 0.589 & 0.586 & 0.358 & 0.454 \\
    &                     & $\gprime - \rprime$ & 0.142 & 0.097 & 0.063 & 0.149 \\
    &                     & $\rprime - \iprime$ & 0.094 & 0.105 & 0.045 & 0.096 \\
    &                     & $\iprime - \zprime$ & 0.230 & 0.235 & 0.106 & 0.227 \\
\hline
    & $20 < \rprime < 21$ & $\uprime - \gprime$ & 0.650 & 0.652 & 0.516 & 0.512 \\
    &                     & $\gprime - \rprime$ & 0.244 & 0.160 & 0.130 & 0.241 \\
    &                     & $\rprime - \iprime$ & 0.185 & 0.191 & 0.108 & 0.170 \\
    &                     & $\iprime - \zprime$ & 0.386 & 0.389 & 0.223 & 0.341 \\
\hline \hline
Red & $17 < \rprime < 18$ & $\uprime - \gprime$ & 0.382 & 0.386 & 0.150 & 0.350 \\
    &                     & $\gprime - \rprime$ & 0.057 & 0.066 & 0.015 & 0.055 \\
    &                     & $\rprime - \iprime$ & 0.040 & 0.069 & 0.011 & 0.038 \\
    &                     & $\iprime - \zprime$ & 0.066 & 0.088 & 0.020 & 0.066 \\
\hline
    & $18 < \rprime < 19$ & $\uprime - \gprime$ & 0.472 & 0.474 & 0.268 & 0.424 \\
    &                     & $\gprime - \rprime$ & 0.078 & 0.067 & 0.028 & 0.087 \\
    &                     & $\rprime - \iprime$ & 0.060 & 0.077 & 0.021 & 0.052 \\
    &                     & $\iprime - \zprime$ & 0.100 & 0.112 & 0.037 & 0.097 \\
\hline
    & $19 < \rprime < 20$ & $\uprime - \gprime$ & 0.512 & 0.514 & 0.453 & 0.473 \\
    &                     & $\gprime - \rprime$ & 0.129 & 0.087 & 0.058 & 0.142 \\
    &                     & $\rprime - \iprime$ & 0.077 & 0.091 & 0.035 & 0.077 \\
    &                     & $\iprime - \zprime$ & 0.148 & 0.156 & 0.069 & 0.147 \\
\hline
    & $20 < \rprime < 21$ & $\uprime - \gprime$ & 0.575 & 0.577 & 0.734 & 0.532 \\
    &                     & $\gprime - \rprime$ & 0.253 & 0.138 & 0.145 & 0.234 \\
    &                     & $\rprime - \iprime$ & 0.132 & 0.144 & 0.073 & 0.125 \\
    &                     & $\iprime - \zprime$ & 0.242 & 0.247 & 0.135 & 0.230 \\
\hline \hline \tablebreak
Blue & $17 < \rprime < 18$ & $\uprime - \gprime$ & 0.347 & 0.352 & 0.119 & 0.292 \\
    &                     & $\gprime - \rprime$ & 0.054 & 0.057 & 0.016 & 0.059 \\
    &                     & $\rprime - \iprime$ & 0.043 & 0.058 & 0.014 & 0.044 \\
    &                     & $\iprime - \zprime$ & 0.096 & 0.104 & 0.031 & 0.099 \\
\hline
    & $18 < \rprime < 19$ & $\uprime - \gprime$ & 0.502 & 0.505 & 0.214 & 0.392 \\
    &                     & $\gprime - \rprime$ & 0.085 & 0.069 & 0.030 & 0.093 \\
    &                     & $\rprime - \iprime$ & 0.061 & 0.073 & 0.025 & 0.062 \\
    &                     & $\iprime - \zprime$ & 0.152 & 0.157 & 0.059 & 0.153 \\
\hline
    & $19 < \rprime < 20$ & $\uprime - \gprime$ & 0.586 & 0.588 & 0.349 & 0.453 \\
    &                     & $\gprime - \rprime$ & 0.141 & 0.097 & 0.061 & 0.150 \\
    &                     & $\rprime - \iprime$ & 0.096 & 0.105 & 0.047 & 0.097 \\
    &                     & $\iprime - \zprime$ & 0.243 & 0.247 & 0.112 & 0.235 \\
\hline
    & $20 < \rprime < 21$ & $\uprime - \gprime$ & 0.648 & 0.653 & 0.512 & 0.513 \\
    &                     & $\gprime - \rprime$ & 0.243 & 0.165 & 0.129 & 0.242 \\
    &                     & $\rprime - \iprime$ & 0.167 & 0.175 & 0.098 & 0.171 \\
    &                     & $\iprime - \zprime$ & 0.383 & 0.384 & 0.221 & 0.346 \\
\hline \hline
\enddata
\label{tab:color_error_cmodel_gal}
\end{deluxetable}

\clearpage

\begin{deluxetable}{ccccccc}
\tablecolumns{7}
\tablewidth{0pt}
\tablecaption{Same as Table~\ref{tab:color_error_model_gal} but for stars
using {\cmodel}.}
\tablehead{
\colhead{Color Cut} & \colhead{Magnitude Limit} & \colhead{Color} &
\colhead{Proper Error} & \colhead{Naive Error} & \colhead{{\tt photo} Error} &
\colhead{Observed Error} 
}
\startdata
All & $17 < \rprime < 18$ & $\uprime - \gprime$ & 0.101 & 0.114 & 0.078 & 0.176 \\
    &                     & $\gprime - \rprime$ & 0.031 & 0.034 & 0.013 & 0.023 \\
    &                     & $\rprime - \iprime$ & 0.024 & 0.034 & 0.009 & 0.024 \\
    &                     & $\iprime - \zprime$ & 0.029 & 0.036 & 0.014 & 0.031 \\
\hline
    & $18 < \rprime < 19$ & $\uprime - \gprime$ & 0.196 & 0.207 & 0.159 & 0.293 \\
    &                     & $\gprime - \rprime$ & 0.034 & 0.039 & 0.020 & 0.029 \\
    &                     & $\rprime - \iprime$ & 0.028 & 0.038 & 0.015 & 0.028 \\
    &                     & $\iprime - \zprime$ & 0.035 & 0.039 & 0.022 & 0.040 \\
\hline
    & $19 < \rprime < 20$ & $\uprime - \gprime$ & 0.291 & 0.306 & 0.243 & 0.328 \\
    &                     & $\gprime - \rprime$ & 0.047 & 0.051 & 0.034 & 0.046 \\
    &                     & $\rprime - \iprime$ & 0.035 & 0.047 & 0.026 & 0.038 \\
    &                     & $\iprime - \zprime$ & 0.057 & 0.061 & 0.047 & 0.085 \\
\hline
    & $20 < \rprime < 21$ & $\uprime - \gprime$ & 0.450 & 0.462 & 0.375 & 0.359 \\
    &                     & $\gprime - \rprime$ & 0.073 & 0.083 & 0.062 & 0.085 \\
    &                     & $\rprime - \iprime$ & 0.065 & 0.082 & 0.057 & 0.071 \\
    &                     & $\iprime - \zprime$ & 0.157 & 0.161 & 0.130 & 0.220 \\
\hline \hline
Red & $17 < \rprime < 18$ & $\uprime - \gprime$ & 0.146 & 0.148 & 0.120 & 0.200 \\
    &                     & $\gprime - \rprime$ & 0.026 & 0.032 & 0.012 & 0.023 \\
    &                     & $\rprime - \iprime$ & 0.022 & 0.033 & 0.009 & 0.023 \\
    &                     & $\iprime - \zprime$ & 0.028 & 0.034 & 0.013 & 0.029 \\
\hline
    & $18 < \rprime < 19$ & $\uprime - \gprime$ & 0.319 & 0.319 & 0.271 & 0.342 \\
    &                     & $\gprime - \rprime$ & 0.029 & 0.032 & 0.020 & 0.029 \\
    &                     & $\rprime - \iprime$ & 0.025 & 0.033 & 0.015 & 0.025 \\
    &                     & $\iprime - \zprime$ & 0.032 & 0.037 & 0.021 & 0.035 \\
\hline
    & $19 < \rprime < 20$ & $\uprime - \gprime$ & 0.470 & 0.470 & 0.470 & 0.444 \\
    &                     & $\gprime - \rprime$ & 0.043 & 0.037 & 0.038 & 0.047 \\
    &                     & $\rprime - \iprime$ & 0.030 & 0.036 & 0.023 & 0.032 \\
    &                     & $\iprime - \zprime$ & 0.041 & 0.043 & 0.033 & 0.052 \\
\hline
    & $20 < \rprime < 21$ & $\uprime - \gprime$ & 0.555 & 0.555 & 0.785 & 0.513 \\
    &                     & $\gprime - \rprime$ & 0.095 & 0.057 & 0.089 & 0.106 \\
    &                     & $\rprime - \iprime$ & 0.049 & 0.053 & 0.044 & 0.051 \\
    &                     & $\iprime - \zprime$ & 0.063 & 0.066 & 0.056 & 0.088 \\
\hline \hline \tablebreak
Blue & $17 < \rprime < 18$ & $\uprime - \gprime$ & 0.111 & 0.121 & 0.070 & 0.056 \\
    &                     & $\gprime - \rprime$ & 0.029 & 0.042 & 0.010 & 0.022 \\
    &                     & $\rprime - \iprime$ & 0.024 & 0.041 & 0.010 & 0.026 \\
    &                     & $\iprime - \zprime$ & 0.033 & 0.041 & 0.018 & 0.035 \\
\hline
    & $18 < \rprime < 19$ & $\uprime - \gprime$ & 0.118 & 0.134 & 0.090 & 0.108 \\
    &                     & $\gprime - \rprime$ & 0.031 & 0.046 & 0.016 & 0.028 \\
    &                     & $\rprime - \iprime$ & 0.029 & 0.046 & 0.017 & 0.030\\
    &                     & $\iprime - \zprime$ & 0.047 & 0.056 & 0.036 & 0.051 \\
\hline
    & $19 < \rprime < 20$ & $\uprime - \gprime$ & 0.293 & 0.307 & 0.230 & 0.244 \\
    &                     & $\gprime - \rprime$ & 0.045 & 0.058 & 0.032 & 0.047 \\
    &                     & $\rprime - \iprime$ & 0.038 & 0.057 & 0.030 & 0.041 \\
    &                     & $\iprime - \zprime$ & 0.079 & 0.085 & 0.067 & 0.101 \\
\hline
    & $20 < \rprime < 21$ & $\uprime - \gprime$ & 0.376 & 0.390 & 0.316 & 0.350 \\
    &                     & $\gprime - \rprime$ & 0.071 & 0.082 & 0.062 & 0.083 \\
    &                     & $\rprime - \iprime$ & 0.064 & 0.080 & 0.058 & 0.072 \\
    &                     & $\iprime - \zprime$ & 0.170 & 0.174 & 0.141 & 0.226 \\
\hline \hline
\enddata
\label{tab:color_error_cmodel_star}
\end{deluxetable}

\clearpage

\begin{deluxetable}{ccccccc}
\tablecolumns{7}
\tablewidth{0pt}
\tablecaption{Same as Table~\ref{tab:color_error_model_gal} but for galaxies
using {\petro}.}
\tablehead{
\colhead{Color Cut} & \colhead{Magnitude Limit} & \colhead{Color} &
\colhead{Proper Error} & \colhead{Naive Error} & \colhead{{\tt photo} Error} &
\colhead{Observed Error} 
}
\startdata
All & $17 < \rprime < 18$ & $\uprime - \gprime$ & 0.231 & 0.239 & 0.203 & 0.263 \\
    &                     & $\gprime - \rprime$ & 0.046 & 0.063 & 0.031 & 0.047 \\
    &                     & $\rprime - \iprime$ & 0.035 & 0.066 & 0.026 & 0.033 \\
    &                     & $\iprime - \zprime$ & 0.070 & 0.090 & 0.051 & 0.079 \\
\hline
    & $18 < \rprime < 19$ & $\uprime - \gprime$ & 0.313 & 0.319 & 0.312 & 0.347 \\
    &                     & $\gprime - \rprime$ & 0.070 & 0.067 & 0.056 & 0.079 \\
    &                     & $\rprime - \iprime$ & 0.052 & 0.070 & 0.048 & 0.054 \\
    &                     & $\iprime - \zprime$ & 0.117 & 0.127 & 0.102 & 0.134 \\
\hline
    & $19 < \rprime < 20$ & $\uprime - \gprime$ & 0.394 & 0.398 & 0.442 & 0.415 \\
    &                     & $\gprime - \rprime$ & 0.120 & 0.098 & 0.112 & 0.144 \\
    &                     & $\rprime - \iprime$ & 0.091 & 0.104 & 0.094 & 0.100 \\
    &                     & $\iprime - \zprime$ & 0.206 & 0.212 & 0.197 & 0.236 \\
\hline
    & $20 < \rprime < 21$ & $\uprime - \gprime$ & 0.477 & 0.483 & 0.607 & 0.489 \\
    &                     & $\gprime - \rprime$ & 0.215 & 0.177 & 0.217 & 0.249 \\
    &                     & $\rprime - \iprime$ & 0.178 & 0.190 & 0.187 & 0.200 \\
    &                     & $\iprime - \zprime$ & 0.340 & 0.346 & 0.352 & 0.373 \\
\hline \hline
Red & $17 < \rprime < 18$ & $\uprime - \gprime$ & 0.262 & 0.268 & 0.245 & 0.286 \\
    &                     & $\gprime - \rprime$ & 0.046 & 0.064 & 0.033 & 0.048 \\
    &                     & $\rprime - \iprime$ & 0.034 & 0.067 & 0.026 & 0.032 \\
    &                     & $\iprime - \zprime$ & 0.063 & 0.086 & 0.046 & 0.067 \\
\hline
    & $18 < \rprime < 19$ & $\uprime - \gprime$ & 0.362 & 0.365 & 0.386 & 0.385 \\
    &                     & $\gprime - \rprime$ & 0.071 & 0.067 & 0.063 & 0.081 \\
    &                     & $\rprime - \iprime$ & 0.049 & 0.069 & 0.045 & 0.048 \\
    &                     & $\iprime - \zprime$ & 0.100 & 0.112 & 0.088 & 0.108 \\
\hline
    & $19 < \rprime < 20$ & $\uprime - \gprime$ & 0.443 & 0.446 & 0.573 & 0.463 \\
    &                     & $\gprime - \rprime$ & 0.127 & 0.094 & 0.124 & 0.150 \\
    &                     & $\rprime - \iprime$ & 0.079 & 0.094 & 0.084 & 0.083 \\
    &                     & $\iprime - \zprime$ & 0.169 & 0.175 & 0.165 & 0.184 \\
\hline
    & $20 < \rprime < 21$ & $\uprime - \gprime$ & 0.518 & 0.523 & 0.889 & 0.539 \\
    &                     & $\gprime - \rprime$ & 0.227 & 0.156 & 0.233 & 0.264 \\
    &                     & $\rprime - \iprime$ & 0.141 & 0.155 & 0.152 & 0.154 \\
    &                     & $\iprime - \zprime$ & 0.269 & 0.275 & 0.274 & 0.297 \\
\hline \hline \tablebreak
Blue & $17 < \rprime < 18$ & $\uprime - \gprime$ & 0.179 & 0.189 & 0.136 & 0.181 \\
    &                     & $\gprime - \rprime$ & 0.042 & 0.060 & 0.027 & 0.045 \\
    &                     & $\rprime - \iprime$ & 0.038 & 0.061 & 0.028 & 0.038 \\
    &                     & $\iprime - \zprime$ & 0.098 & 0.108 & 0.073 & 0.106 \\
\hline
    & $18 < \rprime < 19$ & $\uprime - \gprime$ & 0.259 & 0.266 & 0.236 & 0.271 \\
    &                     & $\gprime - \rprime$ & 0.067 & 0.070 & 0.052 & 0.077 \\
    &                     & $\rprime - \iprime$ & 0.061 & 0.076 & 0.054 & 0.063 \\
    &                     & $\iprime - \zprime$ & 0.157 & 0.163 & 0.135 & 0.172 \\
\hline
    & $19 < \rprime < 20$ & $\uprime - \gprime$ & 0.359 & 0.364 & 0.374 & 0.373 \\
    &                     & $\gprime - \rprime$ & 0.118 & 0.104 & 0.107 & 0.138 \\
    &                     & $\rprime - \iprime$ & 0.107 & 0.117 & 0.106 & 0.114 \\
    &                     & $\iprime - \zprime$ & 0.250 & 0.255 & 0.236 & 0.274 \\
\hline
    & $20 < \rprime < 21$ & $\uprime - \gprime$ & 0.458 & 0.465 & 0.556 & 0.474 \\
    &                     & $\gprime - \rprime$ & 0.213 & 0.183 & 0.213 & 0.247 \\
    &                     & $\rprime - \iprime$ & 0.192 & 0.200 & 0.197 & 0.213 \\
    &                     & $\iprime - \zprime$ & 0.368 & 0.370 & 0.381 & 0.395 \\
\hline \hline
\enddata
\label{tab:color_error_petro_gal}
\end{deluxetable}

\clearpage

\begin{deluxetable}{ccccccc}
\tablecolumns{7}
\tablewidth{0pt}
\tablecaption{Same as Table~\ref{tab:color_error_model_gal} but for stars
using {\psf}.}
\tablehead{
\colhead{Color Cut} & \colhead{Magnitude Limit} & \colhead{Color} &
\colhead{Proper Error} & \colhead{Naive Error} & \colhead{{\tt photo} Error} &
\colhead{Observed Error} 
}
\startdata
All & $17 < \rprime < 18$ & $\uprime - \gprime$ & 0.074 & 0.086 & 0.071 & 0.133 \\
    &                     & $\gprime - \rprime$ & 0.035 & 0.037 & 0.031 & 0.030 \\
    &                     & $\rprime - \iprime$ & 0.029 & 0.038 & 0.023 & 0.029 \\
    &                     & $\iprime - \zprime$ & 0.036 & 0.039 & 0.029 & 0.035 \\
\hline
    & $18 < \rprime < 19$ & $\uprime - \gprime$ & 0.135 & 0.150 & 0.131 & 0.264 \\
    &                     & $\gprime - \rprime$ & 0.038 & 0.040 & 0.035 & 0.035 \\
    &                     & $\rprime - \iprime$ & 0.032 & 0.040 & 0.026 & 0.032 \\
    &                     & $\iprime - \zprime$ & 0.040 & 0.042 & 0.033 & 0.043 \\
\hline
    & $19 < \rprime < 20$ & $\uprime - \gprime$ & 0.203 & 0.230 & 0.214 & 0.366 \\
    &                     & $\gprime - \rprime$ & 0.053 & 0.048 & 0.048 & 0.052 \\
    &                     & $\rprime - \iprime$ & 0.038 & 0.046 & 0.033 & 0.039 \\
    &                     & $\iprime - \zprime$ & 0.048 & 0.050 & 0.043 & 0.062 \\
\hline
    & $20 < \rprime < 21$ & $\uprime - \gprime$ & 0.293 & 0.317 & 0.336 & 0.419 \\
    &                     & $\gprime - \rprime$ & 0.080 & 0.070 & 0.074 & 0.093 \\
    &                     & $\rprime - \iprime$ & 0.055 & 0.064 & 0.051 & 0.060 \\
    &                     & $\iprime - \zprime$ & 0.074 & 0.076 & 0.070 & 0.128 \\
\hline \hline
Red & $17 < \rprime < 18$ & $\uprime - \gprime$ & 0.095 & 0.096 & 0.090 & 0.148 \\
    &                     & $\gprime - \rprime$ & 0.032 & 0.036 & 0.028 & 0.030 \\
    &                     & $\rprime - \iprime$ & 0.028 & 0.037 & 0.024 & 0.029 \\
    &                     & $\iprime - \zprime$ & 0.035 & 0.039 & 0.030 & 0.034 \\
\hline
    & $18 < \rprime < 19$ & $\uprime - \gprime$ & 0.199 & 0.200 & 0.194 & 0.296 \\
    &                     & $\gprime - \rprime$ & 0.036 & 0.037 & 0.033 & 0.036 \\
    &                     & $\rprime - \iprime$ & 0.030 & 0.037 & 0.027 & 0.030 \\
    &                     & $\iprime - \zprime$ & 0.036 & 0.040 & 0.032 & 0.038 \\
\hline 
    & $19 < \rprime < 20$ & $\uprime - \gprime$ & 0.374 & 0.375 & 0.428 & 0.426 \\
    &                     & $\gprime - \rprime$ & 0.051 & 0.041 & 0.048 & 0.055 \\
    &                     & $\rprime - \iprime$ & 0.036 & 0.041 & 0.033 & 0.038 \\
    &                     & $\iprime - \zprime$ & 0.044 & 0.047 & 0.040 & 0.052 \\
\hline
    & $20 < \rprime < 21$ & $\uprime - \gprime$ & 0.518 & 0.518 & 0.698 & 0.499 \\
    &                     & $\gprime - \rprime$ & 0.092 & 0.061 & 0.090 & 0.104 \\
    &                     & $\rprime - \iprime$ & 0.051 & 0.055 & 0.049 & 0.055 \\
    &                     & $\iprime - \zprime$ & 0.061 & 0.062 & 0.058 & 0.085 \\
\hline \hline \tablebreak
Blue & $17 < \rprime < 18$ & $\uprime - \gprime$ & 0.050 & 0.069 & 0.043 & 0.048 \\
    &                     & $\gprime - \rprime$ & 0.034 & 0.045 & 0.028 & 0.030 \\
    &                     & $\rprime - \iprime$ & 0.029 & 0.044 & 0.023 & 0.032 \\
    &                     & $\iprime - \zprime$ & 0.036 & 0.042 & 0.028 & 0.039 \\
\hline
    & $18 < \rprime < 19$ & $\uprime - \gprime$ & 0.072 & 0.095 & 0.062 & 0.070 \\
    &                     & $\gprime - \rprime$ & 0.036 & 0.049 & 0.029 & 0.033 \\
    &                     & $\rprime - \iprime$ & 0.032 & 0.049 & 0.026 & 0.035 \\
    &                     & $\iprime - \zprime$ & 0.046 & 0.054 & 0.040 & 0.050 \\
\hline
    & $19 < \rprime < 20$ & $\uprime - \gprime$ & 0.111 & 0.145 & 0.097 & 0.121 \\
    &                     & $\gprime - \rprime$ & 0.044 & 0.064 & 0.036 & 0.043 \\
    &                     & $\rprime - \iprime$ & 0.040 & 0.064 & 0.036 & 0.042 \\
    &                     & $\iprime - \zprime$ & 0.078 & 0.088 & 0.074 & 0.087 \\
\hline
    & $20 < \rprime < 21$ & $\uprime - \gprime$ & 0.189 & 0.225 & 0.176 & 0.228 \\
    &                     & $\gprime - \rprime$ & 0.062 & 0.090 & 0.056 & 0.066 \\
    &                     & $\rprime - \iprime$ & 0.064 & 0.092 & 0.061 & 0.069 \\
    &                     & $\iprime - \zprime$ & 0.161 & 0.171 & 0.154 & 0.188 \\
\hline \hline
\enddata
\label{tab:color_error_psf_star}
\end{deluxetable}

\clearpage

\begin{deluxetable}{cccccccc}
\tablecolumns{8}
\tablewidth{0pt}
\tablecaption{Same as Table~\ref{tab:chi2_model}, but for variable and 
non-variable stellar objects using {\model}.}
\tablehead{
\colhead{Object Type} & \colhead{Magnitude Limit} & \colhead{Color Cut} &
\colhead{$\uprime$} & \colhead{$\gprime$} & \colhead{$\rprime$} &
\colhead{$\iprime$} & \colhead{$\zprime$} 
}
\startdata
Variable Star & $17 < \rprime < 18$ & All & 7.64 & 103 & 150 & 121 & 20.5 \\
         &                     & Red & 1.67 & 37.4 & 123 & 118 & 21.2 \\
         &                     & Blue & 25.6 & 307 & 234 & 142 & 15.5 \\
\hline
         & $18 < \rprime < 19$ & All & 4.98 & 35.7 & 45.6 & 43.7 & 7.54 \\
         &                     & Red & 1.03 & 8.46 & 31.9 & 40.0 & 6.90 \\
         &                     & Blue & 13.8 & 99.9 & 78.4 & 48.5 & 5.22 \\
\hline
         & $19 < \rprime < 20$ & All & 3.67 & 19.2 & 15.9 & 13.3 & 2.37 \\
         &                     & Red & 0.65 & 2.45 & 6.55 & 10.4 & 2.38 \\
         &                     & Blue & 8.00 & 50.6 & 32.2 & 17.7 & 2.29 \\
\hline
         & $20 < \rprime < 21$ & All & 1.86 & 8.97 & 6.17 & 4.87 & 1.45 \\
         &                     & Red & 0.44 & 1.24 & 2.05 & 3.38 & 1.50 \\
         &                     & Blue & 3.95 & 24.3 & 14.3 & 7.70 & 1.49 \\
\hline \hline
Non-variable Star & $17 < \rprime < 18$ & All & 1.16 & 5.24 & 4.53 & 8.34 & 4.07 \\
         &                     & Red & 1.05 & 4.94 & 4.59 & 8.35 & 4.30 \\
         &                     & Blue & 1.56 & 6.96 & 4.38 & 7.29 & 2.51 \\
\hline
         & $18 < \rprime < 19$ & All & 1.02 & 2.79 & 2.66 & 4.72 & 2.39 \\
         &                     & Red & 0.97 & 2.48 & 2.64 & 4.93 & 2.64 \\
         &                     & Blue & 1.19 & 4.29 & 2.69 & 4.14 & 1.48 \\
\hline
         & $19 < \rprime < 20$ & All & 0.79 & 1.74 & 1.78 & 3.11 & 1.58 \\
         &                     & Red & 0.70 & 1.43 & 1.71 & 3.29 & 1.75 \\
         &                     & Blue & 1.04 & 2.71 & 1.96 & 2.23 & 1.09 \\
\hline
         & $20 < \rprime < 21$ & All & 0.70 & 1.36 & 1.44 & 1.97 & 1.22 \\
         &                     & Red & 0.49 & 1.11 & 1.37 & 2.23 & 1.32 \\
         &                     & Blue & 1.03 & 1.91 & 1.59 & 1.47 & 1.05 \\
\enddata
\label{tab:chi2_vary_star_model}
\end{deluxetable}

\clearpage

\begin{deluxetable}{cccccccc}
\tablecolumns{8}
\tablewidth{0pt}
\tablecaption{Same as Table~\ref{tab:chi2_model}, but for variable and 
non-variable stellar objects using {\cmodel}.}
\tablehead{
\colhead{Object Type} & \colhead{Magnitude Limit} & \colhead{Color Cut} &
\colhead{$\uprime$} & \colhead{$\gprime$} & \colhead{$\rprime$} &
\colhead{$\iprime$} & \colhead{$\zprime$} 
}
\startdata
Variable Star & $17 < \rprime < 18$ & All & 7.49 & 109 & 139 & 116 & 19.5 \\
         &                     & Red & 1.95 & 33.4 & 105 & 112 & 21.5 \\
         &                     & Blue & 22.1 & 304 & 220 & 133 & 14.3 \\
\hline
         & $18 < \rprime < 19$ & All & 4.53 & 29.6 & 35.8 & 34.5 & 5.90 \\
         &                     & Red & 1.39 & 6.25 & 23.0 & 26.7 & 5.20 \\
         &                     & Blue & 10.3 & 84.2 & 63.0 & 42.9 & 4.41 \\
\hline
         & $19 < \rprime < 20$ & All & 3.93 & 21.3 & 14.2 & 10.2 & 2.04 \\
         &                     & Red & 0.88 & 2.16 & 4.20 & 6.56 & 1.99 \\
         &                     & Blue & 5.17 & 31.5 & 19.4 & 12.1 & 2.11 \\
\hline
         & $20 < \rprime < 21$ & All & 2.39 & 11.3 & 6.27 & 4.48 & 1.64 \\
         &                     & Red & 0.44 & 1.45 & 1.61 & 2.99 & 1.34 \\
         &                     & Blue & 2.51 & 11.7 & 6.77 & 4.35 & 1.64 \\
\hline \hline
Non-variable Star & $17 < \rprime < 18$ & All & 1.41 & 4.38 & 4.26 & 7.53 & 
3.29 \\
         &                     & Red & 1.45 & 4.03 & 4.49 & 7.50 & 3.53 \\
         &                     & Blue & 1.23 & 5.52 & 4.24 & 6.79 & 2.14 \\
\hline
         & $18 < \rprime < 19$ & All & 1.32 & 2.25 & 2.44 & 3.88 & 1.80 \\
         &                     & Red & 1.41 & 1.98 & 2.49 & 3.97 & 1.99 \\
         &                     & Blue & 1.13 & 3.06 & 2.45 & 3.30 & 1.39 \\
\hline
         & $19 < \rprime < 20$ & All & 1.20 & 1.70 & 1.59 & 2.20 & 1.33 \\
         &                     & Red & 1.02 & 1.34 & 1.51 & 2.51 & 1.42 \\
         &                     & Blue & 1.28 & 1.93 & 1.62 & 1.97 & 1.30 \\
\hline
         & $20 < \rprime < 21$ & All & 1.30 & 1.68 & 1.28 & 1.52 & 1.45 \\
         &                     & Red & 0.50 & 1.18 & 1.11 & 1.78 & 1.18 \\
         &                     & Blue & 1.36 & 1.75 & 1.32 & 1.45 & 1.48 \\
\enddata
\label{tab:chi2_vary_star_cmodel}
\end{deluxetable}

\clearpage

\begin{deluxetable}{cccccccc}
\tablecolumns{8}
\tablewidth{0pt}
\tablecaption{Same as Table~\ref{tab:chi2_model}, but for variable and 
non-variable stellar objects using {\psf}.}
\tablehead{
\colhead{Object Type} & \colhead{Magnitude Limit} & \colhead{Color Cut} &
\colhead{$\uprime$} & \colhead{$\gprime$} & \colhead{$\rprime$} &
\colhead{$\iprime$} & \colhead{$\zprime$} 
}
\startdata
Variable Star & $17 < \rprime < 18$ & All & 6.10 & 17.5 & 32.3 & 26.1 & 5.35 \\
         &                     & Red & 1.66 & 10.4 & 26.6 & 23.9 & 4.00 \\
         &                     & Blue & 17.2 & 34.8 & 45.9 & 34.3 & 7.64 \\
\hline
         & $18 < \rprime < 19$ & All & 4.61 & 10.4 & 18.6 & 15.2 & 3.02 \\
         &                     & Red & 1.10 & 3.91 & 14.0 & 13.4 & 2.35 \\
         &                     & Blue & 11.8 & 24.5 & 29.2 & 21.3 & 4.17 \\
\hline
         & $19 < \rprime < 20$ & All & 3.60 & 9.48 & 9.98 & 7.60 & 1.71 \\
         &                     & Red & 0.69 & 1.76 & 4.46 & 5.77 & 1.44 \\
         &                     & Blue & 8.17 & 23.6 & 20.6 & 12.7 & 2.30 \\
\hline
         & $20 < \rprime < 21$ & All & 2.18 & 7.66 & 5.95 & 4.16 & 1.30 \\
         &                     & Red & 0.50 & 1.20 & 1.75 & 2.24 & 1.17 \\
         &                     & Blue & 4.43 & 19.2 & 13.7 & 7.62 & 1.61 \\
\hline \hline
Non-variable Star & $17 < \rprime < 18$ & All & 1.16 & 1.11 & 1.11 & 1.40 & 
1.20 \\
         &                     & Red & 1.10 & 1.10 & 1.13 & 1.38 & 1.22 \\
         &                     & Blue & 1.36 & 1.08 & 1.11 & 1.44 & 1.16 \\
\hline
         & $18 < \rprime < 19$ & All & 1.08 & 1.12 & 1.11 & 1.31 & 1.25 \\
         &                     & Red & 1.05 & 1.11 & 1.12 & 1.28 & 1.25 \\
         &                     & Blue & 1.21 & 1.14 & 1.12 & 1.36 & 1.14 \\
\hline
         & $19 < \rprime < 20$ & All & 0.87 & 1.17 & 1.17 & 1.29 & 1.15 \\
         &                     & Red & 0.77 & 1.11 & 1.15 & 1.33 & 1.16 \\
         &                     & Blue & 1.11 & 1.35 & 1.29 & 1.33 & 1.07 \\
\hline
         & $20 < \rprime < 21$ & All & 0.76 & 1.18 & 1.15 & 1.21 & 1.08 \\
         &                     & Red & 0.55 & 1.07 & 1.09 & 1.21 & 1.08 \\
         &                     & Blue & 1.11 & 1.44 & 1.29 & 1.22 & 1.10 \\
\enddata
\label{tab:chi2_vary_star_psf}
\end{deluxetable}

\clearpage

\begin{deluxetable}{cccccc}
\tablecolumns{6}
\tablewidth{0pt}
\tablecaption{Pipeline to observed scatter translation parameters for 
Equation~\ref{eq:pipeline2scatter} as a function of aperture and object type
for each filter.}
\tablehead{
\colhead{Aperture} & \colhead{Object Type} & \colhead{Filter} &
\colhead{$m_0$} & \colhead{$\alpha$} & \colhead{$\beta$} 
}
\startdata
{\model} & Galaxy & $\uprime$ & 16.80 & -9.81 & 0.71 \\
         &        & $\gprime$ & 19.65 & -12.03 & 1  \\
         &        & $\rprime$ & 20.77 & -10.16 & 1 \\
         &        & $\iprime$ & 20.83 & -10.19 & 1 \\
         &        & $\zprime$ & 18.55 & -14.10 & 1 \\ \hline
{\cmodel} & Galaxy & $\uprime$ & 45.29 & -1.87 & -3.18 \\
         &        & $\gprime$ & 20.36 & -8.51 & 1  \\
         &        & $\rprime$ & 19.72 & -12.06 & 1 \\
         &        & $\iprime$ & 20.05 & -11.21 & 1 \\
         &        & $\zprime$ & 19.84 & -7.72 & 1 \\ \hline
{\petro} & Galaxy & $\uprime$ & 76.64 & -0.80 & -2.11 \\
         &        & $\gprime$ & 17.65 & -16.42 & 1  \\
         &        & $\rprime$ & 17.81 & -21.45 & 1 \\
         &        & $\iprime$ & 17.82 & -23.0 & 1 \\
         &        & $\zprime$ & 17.10 & -20.79 & 1 \\ \hline
{\psf}   &  Star  & $\uprime$ & 20.0 & 0.0 & 0.34 \\
         &        & $\gprime$ & 20.0 & 0.0 & 0.34  \\
         &        & $\rprime$ & 20.0 & 0.0 & 0.45 \\
         &        & $\iprime$ & 20.0 & 0.0 & 0.44 \\
         &        & $\zprime$ & 20.0 & 0.0 & 0.12 \\ \hline
\enddata
\label{tab:pipeline2scatter}
\end{deluxetable}

\clearpage

\begin{deluxetable}{cccccc}
\tablecolumns{6}
\tablewidth{0pt}
\tablecaption{$R_{\rm R}$ and $R_{\rm D}$ for galaxies using {\model}.}
\tablehead{
\colhead{Matrix Type} & \multicolumn{5}{c}{Regression Matrix}
}
\startdata
$R_{\rm R}$ & 1    & 0.10 & 0.11 & 0.11 & 0.08 \\
           & 0.10 & 1    & 0.60 & 0.57 & 0.41 \\
           & 0.11 & 0.60 & 1    & 0.81 & 0.59 \\
           & 0.11 & 0.57 & 0.81 & 1    & 0.63 \\
           & 0.08 & 0.41 & 0.59 & 0.63 & 1    \\\hline
$R_{\rm D}$ & 0    & 0.15 & 0.14 & 0.13 & 0.06 \\
           & 0.15 & 0    & 0.11 & 0.08 & -0.05  \\
           & 0.14 & 0.11 & 0    & -0.05 & -0.15 \\
           & 0.13 & 0.08 & -0.05 & 0    & -0.18 \\
           & 0.06 & -0.05 & -0.15 & -0.18 & 0    \\
\enddata
\label{tab:regression_matrices}
\end{deluxetable}

\clearpage

\begin{deluxetable}{cc|cc}
\tablecolumns{4}
\tablewidth{0pt}
\tablecaption{Elements of $A$ and $r_0$ from Equation~\ref{eq:regression_amp}
  for galaxies using {\model}.}
\tablehead{
  \multicolumn{2}{c}{$r_0$} & \multicolumn{2}{c}{$A$}
}
\startdata
$r_{0,ug}$ & 19.3272  & $A_{ug}$ & -6.91268  \\
$r_{0,ur}$ & 19.3886  & $A_{ur}$ & -5.84337  \\
$r_{0,ui}$ & 19.5959  & $A_{ui}$ & -5.74537  \\
$r_{0,uz}$ & 19.7707  & $A_{uz}$ & -7.28647  \\
$r_{0,gr}$ & 18.1922  & $A_{gr}$ & -1.63462  \\
$r_{0,gi}$ & 18.4808  & $A_{gi}$ & -1.43895  \\
$r_{0,gz}$ & 19.2294  & $A_{gz}$ & -2.93277  \\
$r_{0,ri}$ & 15.2442  & $A_{ri}$ & -0.586426  \\
$r_{0,rz}$ & 17.0559  & $A_{rz}$ & -1.04674  \\
$r_{0,iz}$ & 16.6287  & $A_{rz}$ & -1.21666  \\
\enddata
\label{tab:regression_amp}
\end{deluxetable}

\end{document}